\documentclass[aps,showpacs,eqsecnum]{revtex4}

\usepackage{graphicx}
\usepackage{amsmath}
\usepackage{amssymb}

\usepackage{color}

\def\a{\alpha}
\def\b{\beta}
\def\e{\varepsilon}
\def\d{\delta}
\def\g{\gamma}
\def\m{\mu}
\def\t{\tau}
\def\n{\nu}
\def\o{\omega}
\def\s{\sigma}
\def\G{\Gamma}
\def\GG{{\cal{G}}}

\def\O{\Omega}
\def\Si{\Sigma}
\def\ra{\rightarrow}

\def\pd{\partial}
\def\bk{{\bf k}}
\def\bp{{\bf p}}
\def\bq{{\bf q}}
\def\bQ{{\bf Q}}

\def\be{\begin{equation}}\def\ee{\end{equation}}
\def\bea{\begin{eqnarray}}\def\eea{\end{eqnarray}}
\def\nn{\nonumber}

\def\pref#1{(\ref{#1})}

\newcount\bozza \bozza=0
\ifnum\bozza=1
\newdimen\shift \shift=-2truecm
\def\lb#1{%
{\label{#1}\rlap{\kern\shift{$\scriptstyle#1$}}}}
\else\def\lb#1{\label{#1}} \fi

\begin{document}


\title{Ward identity and optical-conductivity sum rule in the
$d$-density  wave state}

\author{L.~Benfatto$^1$}
\email{lara.benfatto@unifr.ch}

\author{S.G.~Sharapov$^{2}$}
\email{sharapov@bitp.kiev.ua}

\author{N.~Andrenacci$^{3}$}

\author{H.~Beck$^3$}
\email{Hans.Beck@unine.ch}
\affiliation{
$^1$D\'epartment de Physique, Universit\'e de Fribourg, P\'erolles,
CH-1700 Fribourg, Switzerland\\
$^2$Istituto Nazionale per la Fisica della Materia
        (INFM);\\
Institute for Scientific Interchange, via Settimio Severo 65,
I-10133 Torino, Italy\\
$^3$Institut de Physique,
Universit\'e de Neuch\^atel, CH-2000 Neuch\^atel, Switzerland}

\date{\today }

\begin{abstract}

We consider the role of the Ward identity in dealing with
the transport properties of an interacting system forming
a $d$-wave modulated charge-density wave or staggered flux phase.
In particular, we address this issue from the point of
view of the restricted optical-conductivity sum rule. Our aim is
to provide a controlled approximation for the current-current
correlation function which allows us also to determine analytically
the corresponding sum rule. By analyzing the role of the vertex
functions in both the microscopic interacting model and in the
effective mean-field Hamiltonian, we propose a non-standard
low-energy sum-rule for this system. We also discuss the possible
applicability of these results for the description of cuprate
superconductors in the pseudogap regime.

\end{abstract}




\maketitle

\section{Introduction}

In the last years a quite important re-examination of the optical
conductivity of high-$T_c$ superconductors (HTSC) has been performed, due
to the improved experimental resolution. Despite the variety of features
observed in the different families of cuprates, when the integral up to
large frequencies of the optical spectra is concerned a common behavior can
be found \cite{Molegraaf:2002:Science,Santander:2002:PRL,Santander:2004,
Homes:2004:PRB,Boris:2004:Science,Ortolani:2004}. This result
is particularly interesting, because it would allow us to distinguish
between different theoretical scenarios for HTSC, in particular
for the pseudogap phase observed in underdoped compounds.  The
optical spectral weight is defined as the integral of the optical
conductivity in a given direction $i=x,y,z$:
\begin{equation}
\lb{weight}
W_i(\omega_M,T) = \int_{-\omega_M}^{\omega_M} \mbox{Re} \,
\sigma_{ii}(\omega,T)d \omega,
\end{equation}
and can be analyzed as a function of both the temperature $T$ and
the cutoff frequency $\omega_M$.  According to this definition,
the weight $W_i$ includes also the condensate peak at $\omega=0$
which develops in the superconducting (SC) state below $T_c$.
Depending on the cut-off $\omega_M$ the sum rule \pref{weight}
acquires different meanings.  When all the optical transitions are
taken into account, Eq.~\pref{weight} expresses simply the so-called
full f-sum rule \cite{Jaklic:2000:AP,Millis:book,Marel:2003},
relating the optical spectral weight to the total carrier density
$n$,

\begin{equation}
\lb{full.rule}
\int_{-\infty}^{\infty} \mbox{Re} \, \sigma(\omega) d \omega
= \frac{\pi n e^2}{m},
\end{equation}
where $m$ is the bare electronic mass. However,
it is usually assumed that when $\o_M$ is of the order of the
plasma frequency only intraband optical transitions relative to the lowest
conduction band $\e_\bk$ contribute to $W(T)$, so
that one obtains the {\em restricted} or {\em partial sum rule}
\cite{Maldague:1977:PRB,Baeriswyl:1987:PRB,Scalapino:1993:PRB}, which
relates $W_i$ to the average value of the diamagnetic term $\t_{ii}$ (see
Eq.~(\ref{j-tau.def}) below),
\be
\lb{weight.band}
W_i(\o_P,T)\equiv W(T) = \frac{\pi e^2}{V} \langle
\tau_{ii} \rangle=
\frac{\pi e^2}{V N} \sum_{\mathbf{k},\sigma}
\frac{\partial^2 \e_\bk }{\partial k_i^2} n_{\bk,\sigma}
\ee
where $n_{\bk,\sigma}$ is the momentum occupation number, $V$ is the
unit-cell volume, $N$ is the number of unit cells, $e$ is the electron
charge, and we set $\hbar = c =1$. In the 2D case $V = a^2$, and in the
quasi-2D case $V = a^2 s$, where $a$ is the lattice spacing and $s$ is the
distance between the layers. In the following we will consider mainly
in-plane processes and isotropic systems, where $W_x=W_y=W$.

The main difference between the restricted and full sum rule is that while
$W(\o_M \to \infty,T)$ is a constant, $W(T)$ given by
Eq.~(\ref{weight.band}) is in general a function of temperature, which
provides information about the interactions between the electrons in the
system. In particular, in a 2D lattice model with a nearest-neighbors
tight-binding dispersion $\e_\bk = - 2t (\cos k_x a + \cos k_y a)$ the
spectral weight Eq.~\pref{weight.band} is proportional to the mean kinetic
energy of the system, $W(T)=- \frac{\pi e^2}{V} \frac{\langle K
\rangle}{2}$. In the absence of interactions $n_{{\bf k} \sigma}
=f(\xi_\bk)$, where $\xi_\bk=\e_\bk-\mu$, $\mu$ is the chemical potential,
and $f(x)$ is the Fermi function. In this case the main temperature
dependence of the spectral weight \pref{weight.band} comes from the
temperature smearing of the Fermi function, and can be easily evaluated
using the Sommerfeld expansion: 
\be 
\lb{weight.normal} 
\frac{W(T)}{(\pi e^2
a^2/V)}=-\frac{1}{N}\sum_\bk \e_\bk f(\xi_\bk)= -\int d \e N(\e)
f(\e-\mu)=\frac{W(0)}{(\pi e^2 a^2/V)}-\frac{\pi^2}{6}c(\mu)T^2 
\ee 
where $N(\e)$ is the density of states for the tight-binding dispersion and
$c(\e)= \e N'(\e)+N(\e)$. By making a quadratic approximation for the
two-dimensional tight-binding band dispersion one would find $c(\mu)=1/4\pi
t$, which is also a good estimate of the exact value obtained using the
true band dispersion and by doping the
system away from half-filling (see also Appendix A). However, for an
interacting system $n_{\bk\s}$ can acquire in general a different
temperature dependence, which influences also $W(T)$.  An example is
provided by the case of a SC instability.  Indeed, according to the BCS
theory, in the SC state the occupation number becomes

\be
\lb{occupation.SC} n_{\bk\s}=[1 - \xi_\bk/E_\bk^{SC} \tanh
(E^{SC}_\bk/2T)], 
\ee 
where $\Delta_\bk$ is the SC gap and $E^{SC}_\bk = \sqrt{\xi^2_\bk +
\Delta^2_\bk}$ is the quasiparticle dispersion in the SC state, so that the
spectral weight \pref{weight.band} decreases below $T_c$. When $W(T)$
corresponds to the kinetic energy this result is understood as the increase
of $<K>$ below $T_c$ due to the particle-hole mixing in the SC state.

These general expectations about the behavior of the restricted optical sum
rule were not confirmed, within several respects, in the experiments on
HTSC. Early measurements of the c-axis spectral weight up to frequencies of
the order of the plasma edge, $\o_P\varpropto 10^4$ cm$^{-1}$, showed that
in YB$_2$Cu$_3$O$_{6+\delta}$ (YBCO) compounds $W_z(T)$ exhibits a quite
anomalous temperature dependence, with a decrease below the pseudogap
temperature, followed by an increase below $T_c$ \cite{c-axis}. Such a
behavior was indeed attributed to the effect of pseudogap opening, combined
with the tunneling character of the transport along the c-axis
direction.

Recently more attention has been instead devoted to the issue of the
spectral-weight behavior for the in-plane optical conductivity, which is a
better probe of the degrees of freedom mostly responsible for the
properties of HTSC. The measurements were performed in
Bi$_2$Sr$_2$CaCu$_2$O$_{8+\delta}$ (BSCCO)
\cite{Molegraaf:2002:Science,Santander:2002:PRL,Santander:2004}, YBCO
\cite{Homes:2004:PRB,Boris:2004:Science} and La$_{2-x}$Sr$_x$Cu$_2$O$_{4}$
(LSCO) \cite{Ortolani:2004} compounds at various cut-off frequencies $\o_M$
between $1000$ cm$^{-1}$ ($0.12$ eV) and $20 000$ cm$^{-1}$ ($2.5$ eV).  A
first issue is the behavior of $W(T)$ below $T_c$. While early measurements
in BSCCO samples show that there is an even faster {\em increase\/} of
$W(T)$ below $T_c$ \cite{Molegraaf:2002:Science, Santander:2002:PRL},
contrary to the prediction of the BCS theory, more recent results in BSCCO
\cite{Santander:2004} show that there is a flattening of $W(T)$ in
underdoped samples for $\omega_M = 8000 \mbox{cm}^{-1}$, while a BCS
behavior below $T_c$ is seen in the overdoped BSCCO and in YBCO samples
\cite{Homes:2004:PRB, Boris:2004:Science}. Also from the theoretical point
of view many proposals arose relative to the problem of the lowering of
in-plane kinetic energy in the SC state
\cite{Hirsch:2000:PRB,Norman:2002:PRB,Eckl:2002,Knigavko:2004}.

Interestingly the behavior of $W(T)$ above the SC transition also shows
unexpected features, which deserve more investigation. Indeed, as
observed in Refs.~\cite{Santander:2002:PRL}, the in-plane optical
sum rule does not show any decrease below the temperature at which
the pseudogap forms, contrary to what found for the c-axis
response. In addition, when the plasma edge is considered as a
cut-off, $W(T)$ shows a ``standard'' $T^2$ temperature dependence,
even though these are clearly strongly-interacting
non-Fermi-liquid systems. However, this result is misleading,
because despite the {\em qualitative} analogy with the free
tight-binding result \pref{weight.normal}, the measured $W(T)$ is
in a strong {\em quantitative} disagreement with the estimate
\pref{weight.normal}. Indeed, as we show in
Appendix~\ref{sec:appendix_A}, the coefficient $c(\mu)$ of
Eq.~\pref{weight.normal} is about one order of magnitude larger
than expected by using a $t$ value estimated by other probes (as
photoemission measurements of the Fermi surface), showing that the
sum rule is far from being conventional already in the normal (non
SC) state \cite{Ortolani:2004}.
Moreover, even faster increase of $W(\o_M,T)$ is
observed at smaller values of $\omega_M$ \cite{Santander:2004,Ortolani:2004}.

For these reasons, the issue that we address in the present paper
is the behavior of the optical-conductivity spectra and sum rule
above $T_c$, but within a model system for the pseudogap state.
Between the several proposals existing in the literature about the
origin of the pseudogap \cite{Loktev:2001:PRP}, we focus in the
present paper on the case where a competing order parameter is
formed before the SC state is established.  In particular, we
refer to the so-called flux phase or $d$-density wave state (DDW)
\cite{Halperin:1968:SSP,Affleck:1988:PRB,Nersesyan:1989:JLTP,Schulz:1989:PRB,
Varma:1997:PRB,Eremin:1998:JETPL,Cappelluti:1999:PRB,
Benfatto:2000:EPJ,Chakravarty:2001:PRB}. We would like to stress
that while a flux phase does not present modulated charge, the
same phenomenological spectrum can be considered as emerging due
to the tendency of the system to form charge order near a quantum
critical point \cite{Castellani}. This scenario was studied in
Ref. \cite{Benfatto:2000:EPJ}, and we will refer in the present
paper also to this point of view, which could be useful in
relating the results presented here not only to cuprates, where
they can be only partly applied, but also to other materials
displaying a true $k$-space modulated CDW (as for example
dichalcogenide materials \cite{Vescoli:1998:PRL,Neto:2001:PRL}).

In a previous publication \cite{Benfatto:2003}, we discussed briefly how a
mean-field description of the DDW state can be compatible with an increase
of the spectral weight below the temperature at which the order parameter
forms. However, this result was not considered from a more general point of
view, which consists in relating the sum rule to the problem of providing a
gauge-invariant approximation for the response functions in a given
microscopic model. As we shall see, the basic requirement of respecting the
charge conservation imposes simultaneously several constraints on the
definition of the current operator, the diamagnetic term and the
corresponding electromagnetic correlation functions. The sum rule then
follows naturally when all these requirements are satisfied within a given
approximation for the microscopic interacting model, and different
approximations can lead to different sum rules. As we shall see, while the
anomalous sum rule derived in Ref.~\cite{Benfatto:2003} can be proposed to
reproduce the experimental data for cuprates, the agreement with the
theoretically obtained optical conductivity is more subtle, and more
detailed features specific of different materials should be considered.  A
more difficult task is to properly define the change of behavior of the sum
rule at different cut-off $\o_M$: this problem is quite general, and while
it is clear that for $\o_M\ra \infty$ the full sum rule \pref{full.rule}
must be recovered, there is as yet no clear understanding of a proper
experimental and theoretical definition of the correct cut-off for the
restricted sum rule in Eq.~\pref{weight.band}. In our case, we shall
discuss how the various restricted sum rules should be realized at
different energy scales, even though an exact result cannot be obtained in
this respect.

The structure of the paper is the following.  We begin by presenting in
Sec.~\ref{sec:sum-rule-Hubbard} the general formalism which is needed to
analyze the optical-conductivity sum rule in an interacting system.  In
Sec.~\ref{sec:DDW-violation} we explicitly study the case of a DDW state,
and we show that we can derive a good approximation for the low-energy
optical conductivity which is however no more related to a known sum rule.
In Sec.~\ref{sec:reduced} we solve this problem by analyzing directly the
reduced, low-energy DDW model Hamiltonian, and we calculate explicitly the
sum rule and the optical conductivity within the proposed mean-field
approach to the DDW transition. We then discuss in Sec. V the results
obtained and summarize the procedure described in the paper. In
Appendix~\ref{sec:appendix_A} we report the evaluation of the sum-rule
behavior for the non-interacting tight-binding model, to quantify the
discrepancy with the experimental data, and some details about the role of
disorder are presented in Appendix~\ref{sec:appendix_B}.

\section{Sum rule in a model with gauge invariant interaction}
\lb{sec:sum-rule-Hubbard}

Let us start by considering a general Hamiltonian describing
interacting electrons in a two-dimensional lattice:
\be
\lb{Hamiltonian} H = -t\sum_{<ij>}
c_{i\sigma}^{\dagger}c_{j\sigma}-\mu\sum_{i}
c_{i\sigma}^{\dagger}c_{i\sigma} + \sum_{ij,\s\s'}
c_{i\sigma}^{\dagger}c_{i\sigma} V(\mathbf{r}_i -
\mathbf{r}_j)c_{j\sigma'}^{\dagger}c_{j\sigma'}
\ee
where the field operator $c^{\dagger}_{i\sigma}$ creates an electron of
spin $\sigma$ at $\mathbf{r}_i$, $t$ is the hopping parameter,
$\langle i j \rangle$ is the sum over nearest neighbor
sites, $V(\mathbf{r}_i - \mathbf{r}_j)$ is the translationally invariant
electron-electron interaction. When rewritten in reciprocal space, the band
dispersion corresponds to $\varepsilon(\mathbf{k}) = - 2 t (\cos k_x a +
\cos k_y a)$. Throughout the paper units $\hbar = k_B =c =1$ are chosen.

In the DDW state a particle-hole coupling is considered at the
characteristic wave-vector $\mathbf{Q} = (\pi/a, \pi/a)$. The notation is
then simplified by halving the Brillouin zone and introducing
two-component electron operators (the DDW equivalent of Nambu
spinors \cite{Nambu:1960:PR})
\be
\lb{spinors}
\chi_{\bk\s}=
\left( \begin{array}{c}
c_{\bk\s} \\
c_{\bk+\mathbf{Q},\s}
\end{array} \right), \qquad
\chi_{\bk \s}^{\dagger} = \left( \begin{array}{cc}
c_{\bk \s}^{\dagger}
\quad c_{\bk + \mathbf{Q},\s}^{\dagger}
\end{array} \right),
\ee
where $c_{\bk \s}^{\dagger}$ and $c_{\bk \s}$
are the Fourier transforms of $c_{i\s}^{\dagger}$
and $c_{i\s}$. The Hamiltonian (\ref{Hamiltonian})
written in terms of $\chi$ becomes
\begin{equation}
\lb{Hamiltonian.spinor}
\begin{split}
H  = &  \sum_{\mathbf{k}, \s}^{\mathrm{RBZ}}
\chi_{\bk \s}^{\dagger} \left[
\frac{1}{2}(\varepsilon_\mathbf{k} + \varepsilon_{\mathbf{k} + \mathbf{Q}}) -
\mu
+ \frac{1}{2}(\varepsilon_\mathbf{k} - \varepsilon_{\mathbf{k} + \mathbf{Q}})
\sigma_3\right]  \chi_{\bk \s} \\
& + \frac{1}{N } \sum_{{\bq}}^{\mathrm{BZ}}
V(\bq)
\sum_{{\bk,\s}}^{\mathrm{RBZ}}
\chi_{\mathbf{k}+ \mathbf{q},\s}^{\dagger} \chi_{\bk \s}
\sum_{{\bp,\s'}}^{\mathrm{RBZ}}
\chi_{\mathbf{p}- \mathbf{q},\s^\prime}^{\dagger}
\chi_{\mathbf{p} \s^\prime},
\end{split}
\end{equation}
where $V(\mathbf{q})$ is the Fourier transform of the potential,
$\sigma_3$ is the Pauli matrix and the sums are over the reduced
(RBZ) and full Brillouin zone (BZ).

After employing the nesting property
$\varepsilon_{\mathbf{k} + \mathbf{Q}} = -
\varepsilon_{\mathbf{k}}$ the kinetic term of the
Hamiltonian (\ref{Hamiltonian.spinor})
takes a simple form
\begin{equation}
\lb{Hamiltonian.0}
H_{0}  = \sum_{\mathbf{k}, \s}^{\mathrm{RBZ}}
\chi_{\bk \s}^{\dagger} \left[ \varepsilon_{\mathbf{k}} \sigma_3 - \mu
\right] \chi_{\bk \s}.
\end{equation}
Accordingly, the  free electron Green's function reads
\begin{equation}
\lb{Green.common}
G_0^{-1}(\mathbf{k}, i\o_n) =  (i\omega_n + \mu)\s_0 -
\varepsilon_{\mathbf{k}} \sigma_3,
\end{equation}
where $\omega_n =   (2n+1) \pi T $ is the fermionic (odd) Matsubara frequency.
The full Green's function of the system $G(p),\,\,  p = (\mathbf{p}, i \o)$
is given by the Dyson equation
\be
\lb{Dyson}
G^{-1}(p) = G_{0}^{-1}(p) - \Sigma(p),
\ee
where the self-energy $\Sigma(p)$ is evaluated at Hartree-Fock level as
\be
\lb{shf}
\Sigma(p) = \sum_{k} G(k)V(\mathbf{p}-\mathbf{k}) .
\ee
where $\sum_k$ is a short hand notation for
$T/N\sum_{i\o_n}\sum_\bk^{RBZ}$.  Observe that for a time-independent
interaction, as the one considered in Eq.~\pref{Hamiltonian}, the
self-energy \pref{shf} does not depend explicitly on the frequency, but we
will keep for convenience this more general notation in the following.  In
the case of superconductivity, the Hartree-Fock approximation for the
self-energy, equivalent to the Eq. \pref{shf} rewritten in the
particle-particle channel, gives the usual BCS result for the Green's
function \cite{Schrieffer}. In the case of DDW order it corresponds instead
to the mean-field Green's function usually considered in the literature
\cite{Benfatto:2003,Wang:2001:PRL,Zhu:2001:PRL, Kee:2002:PRB,
Yang:2002:PRB, Chakravarty:2002:PRL,Sharapov:2003:PRB,aristov,carbotte}.

\subsection{The electrical conductivity and the conductivity sum rule}
\lb{sec:general-sigma-sum}

The optical conductivity can be calculated from the
electromagnetic response kernel
\be
\lb{em-kernel} K_{\m\n}(\bq,
i\O_m) =-\tau_{\m\m}\d_{\m\n}(1-\d_{\n0})+\Pi_{\m\n}(\bq, i\O_m),
\ee
where $\Pi_{\mu\nu}(\bq, i \O_m)$ is the correlation function
\be
\lb{cur-cur}
\Pi_{\mu\nu}(\bq, i\O_m)= \frac{1}{N}\int_0^\b
d\t e^{i\O_m\t}\langle T_\t j_\m(\bq, \t )j_\n( -\bq, 0 )
\rangle .
\ee
Here $\tau_{i i}$ is the diamagnetic (or stress)
tensor, $\tau$ is imaginary time, $\b = 1/T$, and $\O_m = 2 \pi m
T$ is the bosonic Matsubara frequency. The index $\mu=(i,0 )$ with
$i=1,2$ indicates spatial and time components respectively, so
that the particle current operator $j_\mu(\bq, \t) = (j_i(\bq,
\t), j_0(\bq, \t))$ consists of the particle current density,
$j_i(\bq, \t)$ and the particle density, $j_0 (\bq, \t)$.  As
usual, the particle current and the diamagnetic tensor are defined
as the first and second order derivatives of the Hamiltonian
$H({\bf A})$ in the presence of the vector potential ${\bf A}$
with respect to ${\bf A}$ itself \cite{Scalapino:1993:PRB}:
\be
\lb{j-tau.def} H(A_i)\approx H(0)  - \sum_{j} \left[ e
A_i(\mathbf{r_j}) j_i(\mathbf{r_j}) - \frac{e^2}{2}
A_i^2(\mathbf{r_j}) \tau_{ii}(\mathbf{r_j}) \right],
\ee
so that the total current density is expressed as $J_i (\mathbf{r})= -
\delta H/\delta A_i(\mathbf{r}) = e j_i(\mathbf{r}) - e^2
\tau_{ii}(\mathbf{r}) A_i(\mathbf{r})$, and by evaluating $\langle J_i (q)
\rangle$ within the linear response theory
\cite{Scalapino:1993:PRB,Enz:book,Schrieffer}, one obtains $J_{\mu}(q)=e^2
K_{\mu\nu}(q)A_{\nu}(q)$ with the electromagnetic kernel
(\ref{em-kernel}). Then using that $\mathbf{A}(\o) =
\mathbf{E}(\o)/i(\o+i0)$, where $\mathbf{E}$ is the electric field, one
finally arrives at the Kubo formula
\be
\lb{conductivity.def} \s(\o)=-ie^2\frac{K_{i
i}(\bq=0,\o)}{V(\o+i0)}= ie^2\frac{<\tau_{i i}>-\Pi_{i
i}(\bq=0,\o)}{V(\o+i0)},
\ee
where the standard analytic continuation $i \Omega_m \to \omega +i 0$ was
made. To avoid confusion, along the paper we will indicate the imaginary
bosonic frequencies with $i\O_m$, the imaginary fermionic frequencies with
$i\o_n$ and the real frequencies with $\o$. Since an isotropic system is
considered we can omit the index $i$ as done in the LHS of
Eq.~\pref{conductivity.def} and in what follows.

Taking the real part of \pref{conductivity.def}, one obtains $\mbox{Re}
\s(\o)=(\pi e^2/V) \d(\o)[<\t>-\mbox{Re} \Pi(\mathbf{0},\o)]+ (e^2/V)
\mbox{Im}\Pi(\mathbf{0},\o)/\o$.  In the presence of disorder the
coefficient of the $\d(\o)$ vanishes, so that $\mbox{Re} \Pi(\bq=0,\o\ra
0)=<\t>$ and only the regular part of $\s(\o)$ survives.  As a consequence,
one usually defines the optical conductivity only through the imaginary
part of $\Pi(\bq=0,\omega)$:
\be
\lb{Re.sigma}
\mbox{Re} \s(\o)= \frac{e^2}{V} \frac{\mbox{Im} \Pi(\bq = \mathbf{0},\o)}{\o},
\ee
so that using the Kramers-Kronig (KK) relations for $\Pi(\bq =0,\o)$
one can derive the well-know sum rule:
\be
\lb{common.rule}
W(T)=\int_{-\infty}^{\infty}
\mbox{Re} \sigma(\omega) d \omega = \frac{e^2}{V} \int_{-\infty}^{\infty}
\frac{\mbox{Im} \Pi(\bq = \mathbf{0},\omega)}{\omega}
d \omega = \frac{\pi e^2}{V}
\mbox{Re}\Pi(\bq = \mathbf{0},\o=0)=\frac{\pi e^2}{V} <\tau>.
\ee

The form of $H(\mathbf{A})$ itself depends on the microscopic model and
thus on the way the vector potential ${\bf A}$ enters the Hamiltonian of
the system. When a continuum model is considered instead of
Eq.~\pref{Hamiltonian}, the kinetic term is expressed as $\int
(-\nabla)^2/2m$ and ${\bf A}$ is inserted using the minimal coupling
prescription $- i \nabla \to - i \nabla - e \mathbf{A} $. For lattice
systems the equivalent of the minimal coupling prescription is the
so-called {\em Peierls ansatz}
\cite{Jaklic:2000:AP,Millis:book,Scalapino:1993:PRB}, which corresponds to
insert the gauge field ${\bf A}$ in eq.  \pref{Hamiltonian} by means of the
substitution $c_i\ra c_i e^{- i e \int {\mathbf A}\cdot d {\mathbf r}}$. In
this case, it is clear that when the interaction term of the Hamiltonian is
a density-density interaction, as in Eq.~\pref{Hamiltonian}, only the
kinetic hopping term is modified, while the interaction term is gauge
invariant (GI).  As a result, the current/density operator and the
diamagnetic tensor can be expressed (for small ${\bf q}$) as:

\bea
\lb{defj}
j_\m(\bq,t)&=&\frac{1}{N}\sum_{\mathbf{k}, \sigma}
v_\mu(\bk) c^{\dagger}_{\mathbf{k}- \mathbf{q}/2\s}
c_{\mathbf{k}+\mathbf{q}/2\s}=\frac{1}{N}\sum_{\bk,\s}^{RBZ}
\chi^+_{\bk-\bq/2}\gamma_\mu(\bk-\bq/2,\bk+\bq/2)\chi_{\bk+\bq/2},\quad
\bq\to 0\\
\lb{deft} \tau_{ii}&=&\frac{1}{N}\sum_{\bk,\s} \frac{\partial^2
\e_\bk }{\partial k_i^2} n_{\bk,\sigma} \eea
where
\be \lb{vbare}
v_\mu(\bk)=(v_\bk^F,1),\quad
\gamma_\mu(\bk-\bq/2,\bk+\bq/2)=( v_{\bk}^F\s_3, \s_0), \quad  \bq\to 0\,,
\ee
and $(v_{\bk}^F)_i=\pd \e_\bk/\pd k_i$ is the Fermi velocity
\cite{footnote1}. Note that if a quadratic band dispersion $\e_\bk = \bk^2/2m$
is assumed, the tensor $\tau_{ii}$ reduces to $n/m$, where $n$ is the total
carrier density, so that Eq.~\pref{common.rule} reduces to the f-sum rule
\pref{full.rule}, which is temperature independent. Instead, for a
tight-binding nearest neighbors lattice dispersion, according to the
definition \pref{deft}, $\tau_{ii}$ is proportional to the kinetic energy,
and the sum-rule \pref{weight.normal} is recovered. Observe that formally
the sum rule \pref{common.rule} always requires the
integration up to an infinite cut-off energy. Nevertheless,
an intrinsic finite cut-off energy is provided by the energy scale below
which a given model can be considered as a good approximation for the real
system. As a consequence, while the full f-sum rule is always satisfied at
enough large energy scales, the restricted optical sum rule relative to a
given tight-binding interacting model is expected to hold only below some
intrinsic energy scale, whose definition is not universal.  We would like
to stress that the definitions \pref{defj}-\pref{deft} follow from the
Hamiltonian \pref{Hamiltonian} once that a gauge-invariant form is chosen
for the interaction term. However, this assumption is invalid when for
example ``occupation modulated'' hopping terms are present
\cite{Hirsch:2000:PRB}, or when an ``effective'' interacting model is
considered, in a sense that we will specify below (see Sec. IV).

\subsection{Gauge invariance and the sum rule}

The derivation of the sum rule  presented above is rather formal,
and does not allow one to understand that the sum rule is just a
different way of stating the gauge invariance of the theory. To
gain a deeper insight into the relation between these two aspects,
it is useful to consider here the sum-rule derivation presented in
Ref.~\cite{Enz:book}. The starting point is the observation that
in a GI theory there is a gauge freedom to choose
whether the applied electric field $\mathbf{E} = - \partial_t \mathbf{A} -
\nabla \varphi $ is included in the Hamiltonian \pref{Hamiltonian}
either via the vector potential $\mathbf{A}$ ($\varphi=0$) or by
considering a scalar potential $\varphi$ ($\mathbf{A} =0$).
Obviously, the conductivity derived from two equivalent
Hamiltonians $H(\mathbf{A})$ and $H(\varphi)$ must be the same,
but this is only guaranteed by the charge conservation
\be
\lb{e-conserv} e \partial_t j_0(\mathbf{q},t) + ie \mathbf{q}
\cdot \mathbf{j}(\mathbf{q},t)=0.
\ee
The proof considered in Ref.~\cite{Enz:book} that $\sigma(\omega)$ derived
from $H(\mathbf{A})$ and $H(\varphi)$ are the same is based on the identity
\be \lb{Enz.identity}
\int_{-\infty}^{\infty} d \o\mbox{Re}
\sigma(\o)= \frac{\pi e^2}{V
N} \lim_{q_i \to 0} \frac{1}{q_i} \langle[j_{0}(\mathbf{q},t),
j_i(-\mathbf{q},t)] \rangle,
\ee
which is obtained by
using the charge conservation \pref{e-conserv}.
For example, substituting in Eq.~(\ref{Enz.identity})
$j_0(\mathbf{q},t) = \sum_{\mathbf{k}, \sigma}
c^{\dagger}_{\mathbf{k} - \mathbf{q}/2,\s}
c_{\mathbf{k}+ \mathbf{q}/2,\s}$
and the free-electron expression $\mathbf{j}(\mathbf{q},t) =
(1/m) \sum_{\mathbf{k}, \sigma} \mathbf{k}
c^{\dagger}_{\mathbf{k} - \mathbf{q}/2,\s}
c_{\mathbf{k}+ \mathbf{q}/2,\s}$, corresponding to $\e_\bk=\bk^2/2m$,
returns the full f-sum rule \pref{full.rule}.

Another way to state the relation between the sum rule and the GI
uses instead the generalized electromagnetic kernel
\pref{em-kernel}.  As discussed in Ref. \cite{Schrieffer} with
reference to the SC case, the requirements of charge conservation
($q_\mu J_\mu(q)=0$) and invariance of the theory under the gauge
transformation $A_\mu(q)\ra A_\mu(q)+iq_\mu \Lambda(q)$ are
fulfilled when the condition \be \lb{gen} q_\m
K_{\m\n}(q)=K_{\m\n}(q)q_\nu=0, \qquad q = (\mathbf{q}, \omega)
\ee is satisfied. In particular, the following relation must hold:
\be \lb{GI} \Pi_{ii}(\bq\ra 0,\o=0) = \langle \tau_{ii} \rangle.
\ee This equality states that the diamagnetic term is canceled
out by the {\em static} limit $(\o =0, \bq \to 0)$ of the (real)
part of the current-current bubble, while deriving the
Eq.~\pref{common.rule} we used the relation between the {\em
dynamic} $(\bq =0,\o \to 0)$ limit of the bubble and the stress
tensor. However, in deriving Eq.~\pref{common.rule} we assumed the
presence of disorder, whose role is crucial in restoring the
equality between the static and dynamic limits of the
current-current correlator $\Pi (q)$. Indeed, while in a clean
system the dynamic limit of the bubble vanishes, in the presence
of disorder it coincides with the static limit, which in turn is
equal to the diamagnetic term: $\mbox {Re}\Pi(\o\ra 0,\bq=0)=\mbox
{Re}\Pi(\o=0,\bq\ra 0)=<\tau>$, and then Eq.~\pref{common.rule}
follows.

\subsection{Ward identity and vertex function}
\lb{sec:WI}

The advantage of the derivations \pref{Enz.identity} and \pref{GI} of the
sum rule is that they show explicitly that it has to be regarded as a
consequence of the charge conservation. Moreover, it allows one to see that
once a given approximation is used in evaluating the current-current
correlation function, it also fixes the sum rule that will follow from such
an approximation. However, a quite difficult task is to implement an
approximation for both the Green's function and the current-current
correlator which preserves the condition \pref{gen}, necessary for
maintaining the GI of the theory. In particular, when the
Hartree-Fock self-energy \pref{shf} is used and the bubbles $\Pi_{\mu\nu}$
are evaluated in the lowest-order approximation:
\be
\lb{pmf}
\Pi_{\mu\n}^{(\gamma)}(\bq,i\O_m)=
-2\sum_k \mbox{Tr} [G(\bk-\bq/2,i\o_n+i\O_m)
\g_\mu(\bk-\bq/2,\bk+\bq/2)
G(\bk+\bq/2, i \o_n)\g_\n(\bk+\bq/2,\bk-\bq/2)],
\ee
the GI is {\em not} in general preserved, as it is know for SC and as we
shall see explicitly in Sec.~\ref{sec:reduced} in the case of DDW
(the factor 2 in the previous equation is due to the spin summation).
A general field theoretical approach that solves the difficulties
with charge conservation and gauge invariance, originally present
in the mean-field (bare vertex) formulation of the BCS theory, was
developed by Nambu \cite{Nambu:1960:PR} and discussed in detail in Chapter
8 of \cite{Schrieffer}, so that here we only introduce the main
definitions and stress the points necessary for the consideration
of the DDW state.

As shown in \cite{Schrieffer}, the current-current correlator, defined above
in Eq. \pref{cur-cur}, can be expressed in terms of the full
Green's functions \pref{Dyson}, the  bare vertex $\g_\m$ and the full vertex
function $\G_\nu$ as follows
\be
\lb{pigi}
\Pi_{\mu\n}(\bq,i\O_m)= -2 \sum_k \mbox{Tr}
[G(k_-)\g_\n(\bk_-, \bk_+)
G(k_+)\G_\mu(k_+,k_-) ],
\ee
where $k_+ = ({\bf k_+}, i\o_n+i\O_m,)$, $k_- = ({\bf k_-}, i\o_n,)$ with
${\bf k_\pm}=\bk\pm\bq/2$.
The important property of the current-current correlation function
\pref{pigi} is that the condition \pref{gen} is preserved whenever
the vertex function satisfies the {\it generalized Ward identity} (GWI):
\be
\lb{wi}
q_\mu\G_\mu(p_+,p_-)=G^{-1}(p_-)-G^{-1}(p_+).
\ee
The GWI is nothing but the charge conservation
law \pref{e-conserv} rewritten using the Greens' and vertex functions.  If
the Green's function given by Dyson equation \pref{Dyson} is evaluated within
the Hartree-Fock approximation \pref{shf}, then the vertex function
satisfying the GWI is also the solution of the following integral equation:
\be
\lb{ver}
\G_{\m}(p_+,p_-)=\g_{\m}(\mathbf{p_+},\mathbf{p_-})+\sum_kG(k_+)
\G_\m(k_+,k_-)G(k_-)V(\bp-\bk).
\ee
The analytical solution of Eq.~\pref{ver} cannot be easily determined,
except that in the static limit, when $\G_i$ is given by
\be
\lb{WI-dif}
\G_i(p,p)=
\g_i(\bp,\bp)+\frac{\pd \Si(p)}{\pd \bp_i}=- \frac{\partial
G^{-1}(p)}{\partial {\bf p}_i}= G^{-1}(p)\frac{\partial
G(p)}{\partial {\bf p}_i} G^{-1}(p)
\end{equation}
Indeed, if one puts $q=0$ in Eq. \pref{ver} (which corresponds, as usual,
to the static limit $\o=0, \bq\ra 0$ when analytical continuation
$i \O_m \to \o + i0$ is made),
by means of the previous relation one obtains:
\bea
\G_{i}(p,p)&=&\g_{i}(p,p)\, + \, \sum_{k} G(k) \G_i (k,k) G(k)
V(\bp-\bk)\,= \,\g_{i}(p,p)\, + \, \sum_{k} \frac{\partial G(k)}{\partial
{\bf k}_i} V(\bp-\bk) \nonumber \\
&=&\g_i(p,p)\, - \, \sum_{k} G(k)
\frac{\partial V(\bp-\bk)}{\partial {\bf k}_i}
\,=\,\g_i(p,p)+\frac{\partial}{\partial {\bf p}_i} \sum_{k} G(k)
V(\bp-\bk)\nonumber \\
\lb{gsol}
&=& \g_i(p,p)+\frac{\pd \Si(p)}{\pd
\bp_i}.
\eea
Here we used the fact that the potential $V$ is non-separable, viz. it
depends on the difference $\mathbf{p} - \mathbf{k}$, as it is expected for
a GI density-density interaction. Observe also that this result can be
obtained directly from the GWI \pref{wi} by taking the limit $\o=0,\bq \to
0$. For example, one can easily check that WI \pref{WI-dif} is satisfied for
the free electron Green's function \pref{Green.common} taken together with
the bare vertex (\ref{vbare}). It is worth noting that in the case of SC
the behavior of the vertex function at zero frequency and momentum is
completely different, and indeed $\G_i(p,p)$ is divergent as the inverse of
the phase-mode dispersion \cite{Nambu:1960:PR,Schrieffer}. Indeed, the
equivalent of Eq. \pref{wi} contains for the SC case a combination of
Green's functions and Pauli matrices that cannot be reduced to the
derivative of $G^{-1}$ as in Eq. \pref{WI-dif}. Here, however, the equivalent
of the gapless phase mode is not present, because there is no Goldstone mode
when a discrete symmetry is broken, and $\G_i(p,p)$ turns out to be
finite.

\subsection{Symmetrized expression for $T=0$ dc conductivity}
\lb{sec:sigma-sym}

In practice, since the exact expressions for $G$ and $\Gamma_i$ are
unknown, the consistency of an approximated calculation of the conductivity
can be guaranteed if the {\it approximated} expressions for $G$ and
$\Gamma_i$ satisfy the GWI \pref{wi}.  Observe that what we obtained in
\pref{gsol} is the limit $\o=0,\bq\ra 0$ of $\G$, but in the calculation of
the optical conductivity it is the opposite limit which is needed. However,
at least in the presence of impurities, or at $T=0$, the static and dynamic
limits commute. Unfortunately, a generalization of the result \pref{gsol}
to finite frequency cannot be obtained from the equation \pref{ver} for a
generic potential, by means of, e.g., a perturbative method. Since our
final task is to find an approximation for the optical conductivity which
allows us also to estimate the corresponding sum rule, let us analyze the
utility of the result \pref{gsol}. First, we note that the knowledge of the
vertex function at zero frequency allows one to find an exact result for
the dc conductivity at $T=0$.  To show this it is convenient to think of $2
\times 2$ matrices $A$ and $B$ as being represented by two column vectors
of $2\times 2$ matrix elements and rewrite $\mbox{Tr}$ of the matrix
product as the scalar product:
\be
\lb{def.A*B}
\mbox{Tr} [A B] \equiv \sum_{\a\b} (\vec A)_{\a\b} (\vec B)_{\b\a} =
\vec A\cdot \vec B.
\ee
Accordingly, by introducing the vector
\be
(\GG(k_+,k_-) \vec \G(k_+,k_-))_{\a\b} \equiv \sum_{\g\d}
G_{\a\g}(k_+)G_{\d\b}(k_-)\G_{\g\d}(k_+,k_-),
\ee
we can rewrite
correlation function \pref{pigi} as follows:
\be
\lb{piim}
\Pi_{ij}(q)  =- 2
\int \frac{d^3 k}{(2 \pi)^3}
\GG(k_+,k_-)\vec\G_i(k_+,k_-)\cdot \vec\g_j(\bk_-,\bk_+),
\ee
where since we are considering the $T=0$ case we have an integration
over the real frequency instead of the Matsubara sum and
the argument of the polarization
operator is $q = (\bq, \omega)$.

The dc conductivity is determined by the imaginary part of the derivative
of the correlation function, which in turn is given by:
\be
\lb{pi-derivative}
\begin{split}
& \left. \frac{\partial \Pi_{ij}(\bq =0, \o)}{\partial \o} \right|_{\o=0}\\
& = - 2
\int \frac{d^3 k}{(2 \pi)^3} \left. \left[
\GG^{\prime}_\o(k_+,k_-)\vec\G_i(k_+,k_-)\cdot \vec\g_j(\bk_-,\bk_+)+
\GG(k_+,k_-)\vec\G_{\o i}^{\prime}(k_+,k_-)\cdot \vec\g_j(\bk_-,\bk_+)
\right] \right|_{\o=0, \bq =0},
\end{split}
\ee
where $\GG'_\o, \G'_\o$ indicate the derivative with respect to $\o$.
The expression \pref{pi-derivative} can be further simplified
by using the equation for vertex \pref{ver} taken at $T =0$
and its derivative with respect to $\o$:
\begin{subequations}
\begin{align}
\lb{gamma}
& \vec \g_i(\bk, \bk) = \vec\G_i (k_+,k_-) -
\int \frac{d^3 p}{(2 \pi)^3}
\GG (p_+,p_-)\vec\G_i (p_+,p_-) V(\bk-\bp), \qquad \bq =0\\
& \lb{Gamma-derivative}
\vec\G_{\o i}^{\prime} (k_+,k_-) =\int \frac{d^3 p}{(2 \pi)^3}
[\GG_{\o}^{\prime} (p_+,p_-)\vec\G_i (p_+,p_-) +
\GG (p_+,p_-)\vec\G_{\o i}^{\prime} (p_+,p_-)]V(\bk-\bp).
\end{align}
\end{subequations}
Substituting $\vec \g_i$ from Eq.~\pref{gamma} in Eq.~\pref{piim}
and using \pref{Gamma-derivative} we obtain 
\be
\lb{pi-derivative-final} 
\left. \frac{\partial \Pi_{ij}(\bq =0,
\o)}{\partial \o} \right|_{\o=0} = - 2 \int \frac{d^3 k}{(2
\pi)^3} \left. \GG^{\prime}_\o(k_+,k_-; \bq =0)\vec\G_i(k,k)\cdot
\vec\G_j(k,k) \right|_{\o=0}. 
\ee 
Our derivation is similar to the derivation of the symmetrized expressions
for the derivatives of the polarization operator considered in
Ref.~\cite{Gusynin:1992:JETP}, where also the derivative of Bethe-Salpeter
kernel enters the analog of Eq.~\pref{Gamma-derivative}. The useful
property of the representation (\ref{pi-derivative-final}) for
$\Pi_{\o}^{\prime}$ is that it contains two full vertex functions
$\Gamma_i$. The corresponding expression for the dc conductivity $\s_{dc}$
coincides with the result derived by Langer [Eq.~(4.8) of
Ref.~\cite{Langer:1962a:PR}] in early 60s using a completely different
approach, consisting in introducing a {\em symmetric bubble}
\be 
\lb{pisim} 
\Pi_{ij}^{sym}(i\O_m)=-2\frac{T}{N}
\sum_{\bk,i\o_n}^{RBZ}Tr[G(\bk,i\o_n+i\O_m)\G_i(k,k)
G(\bk,i\o_n)\G_j(k,k)], 
\ee 
obtained by using {\em two} corrected
vertices, evaluated at zero external frequency, and whose
derivative at zero frequency and temperature coincides with the
result \pref{pi-derivative-final}. Then in the limit
$T \to 0$ the leading  term of Langer's expression for the dc conductivity is
obtained from \pref{pisim} via 
\be 
\lb{sigma-dc.symm}
 \s_{dc}  =\lim_{\o \ra 0} \mbox{Re} \s(\o)
= \frac{e^2}{V} \mbox{Im}\left .\pd_\o[\Pi^{sym}_{ii}(i \O_m \to
\o)]\right|_{\o=0}. 
\ee 
In Eq. \pref{gsol} the vertex function with coinciding fermion momenta and
energies, $k_+ = k_- =k$ is related to the self-energy $\Sigma(k)$ by the
WI \pref{WI-dif} \cite{Langer:1962a:PR,Sharapov:2003:PRB}. Thus one can
immediately see that whenever $\Si(k)$ depends on the momentum $\bk$, the
dc conductivity \pref{sigma-dc.symm} would be different from the value
obtained using the bare bubble \pref{pmf}.

From the previous considerations one can argue that, in the
absence of a solution for the vertex function $\G$ at finite
frequency, a better approximation for the conductivity in the DDW
state is provided by the bubble \pref{pisim}, which gives at least
an {\em exact} result for the dc conductivity at $T=0$ (see also
Ref.~\cite{Sharapov:2003:PRB}).  In other words, by evaluating the
symmetric bubbles \pref{pisim} at finite frequency one can still
capture the behavior of $\s(\o)$ at small $\o$.  At the same time,
we do also expect that this assumption will lead to a new result
for the sum rule \pref{common.rule}, because the symmetric bubble
\pref{pisim} is no more connected to the diamagnetic term
\pref{deft} by any relation. However, as we shall see in the next
Section, the sum rule for the bubbles \pref{pisim} can be obtained
analytically by using the analogies between the results discussed
up to now and the properties of the reduced Gaussian model, where
the vertex equation admits the solution Eq.~\pref{gsol} at all
frequencies.

\section{Violation of the GI with the bare vertex in the DDW state}
\lb{sec:DDW-violation}

\subsection{The mean-field DDW Hamiltonian}
\lb{sec:DDW-Hamiltonian}

The previous discussion was generically referred to any system
displaying a particle-hole instability at the wave vector $\bQ$.
However, in the Hartree-Fock approach one usually selects a
particular form for the mean-field Green function $G$ and then
solves the self-consistency equation corresponding to implement
the Dyson equation \pref{shf}.  In the DDW case, one approximates the
general interacting Hamiltonian \pref{Hamiltonian} with the model:
\be
\lb{int}
H_I=-\frac{V_0}{2N}\sum_{{k,k'\atop{\s,\s'}}} w_d(\bk) w_d(\bk') 
c_{\bk\sigma}^\dagger c_{\bk+\bQ\sigma} 
c_{\bk'+\bQ\sigma'}^\dagger c_{\bk'\sigma'},
\ee
where $w_d(\bk)=(\cos k_xa-\cos k_ya)/2$. By defining  $iD_0= -(V_0/N)
\sum_{\bk\s}w_d(\bk)<c_{\bk+\bQ\sigma}^\dagger c_{\bk\sigma}>$ we obtain the
following mean-field DDW Hamiltonian 
\begin{equation}
\lb{DDW-Hamiltonian}
H =  \sum_{\mathbf{k},\sigma} [(\epsilon_\bk -\mu)
c_{\bk\sigma}^\dagger c_{\bk\sigma} +
i D_\bk c_{\bk\sigma}^\dagger c_{\bk+\bQ\sigma}] =
\sum_{\mathbf{k},\sigma}^{RBZ}
\chi_{\bk \s}^{\dagger}
\left[ \varepsilon_\mathbf{k} \sigma_3 -D_\bk \sigma_2- \mu
\right] \chi_{\bk \s},
\end{equation}
where $D_\bk = D_0 w_d(\bk)$ is the gap, known as
the DDW gap \cite{Chakravarty:2001:PRB}, arising from the formation of the
state with circulating currents below a characteristic temperature
$T_{DDW}$ \cite{noteCDW}. This Hamiltonian corresponds to an explicit solution
of Eqs.~\pref{Dyson}-\pref{shf} with
\be
\lb{self-DDW}
\Si(\bk)=-D_\bk \s_2,
\ee
so that the full Green's function \pref{Dyson} reads:
\be
\lb{Green-DDW}
G^{-1}(\mathbf{k}, i\o_n)=  (i\omega_n + \mu)\s_0 -\e_\bk\s_3+D_\bk\s_2.
\ee
The corresponding self-consistency equations for the order parameter $D_0$
and for the chemical potential $\mu$ read:
\bea
\lb{sc1}
& &\frac{2V_0}{N}\sum_\bk^{RBZ}
\frac{w_d^2(\bk)}{E_\bk}[f(\xi_{-,\bk})-f(\xi_{+,\bk})]=1,\\
\lb{sc2} & & \frac{2}{N}\sum_\bk^{RBZ}
[f(\xi_{-,\bk})+f(\xi_{+,\bk})] =n, \eea where
$E_\bk=\sqrt{\e_\bk^2+D_\bk^2}$, and $\xi_{\pm,\bk}=-\mu\pm E_\bk$
are the two excitation branches associated with the formation of
DDW order, which breaks translation symmetry. 
Observe that to obtain the Eq. \pref{sc2} we used the fact that the
occupation number $n_{\bk \s}$ in the DDW is given, according to the 
Green's function \pref{Green-DDW}, by $n_{\bk\s}=(1/2E_\bk) [E_\bk (f(\xi_+)
+f(\xi_-))+\e_\bk (f(\xi_+)-f(\xi_-))]$. This allows us also to evaluate
the diamagnetic term \pref{deft} and the corresponding sum rule as:
\be
\lb{deftmf}
\frac{W(D,T)}{(\pi e^2 a^2/V)}
=<\tau>=-\frac{1}{N}\sum^{RBZ}_\bk
\frac{\e^2}{E}[f(\xi_+)-f(\xi_-)],
\ee
where we used the fact that $(\partial^2 \e_\bk/\partial k_i^2)=+2ta^2
\cos(k_ia)$ and we omitted the explicit dependence of $\e, E, \xi_{\pm}$ on
$\bk$. At this level we have performed an approximation on both the
self-energy and the Green's function of the original, interacting
system. To obtain now a GI approximation for the optical conductivity,
i.e. an approximation which gives Eq. \pref{deftmf} as the
integral of $\s(\o)$, we should also evaluate the vertex function 
\pref{ver}. Indeed, as we show with an explicit calculation in the next
section, the bubble \pref{pmf} with a bare vertex $\gamma$
violates this requirement. In general, if the optical conductivity cannot
be calculated by means of the exact vertex function \pref{ver}, but a
different approximation is used, one cannot expect any more to find Eq.
\pref{deftmf} as the corresponding sum rule, but this has to be calculated
explicitly, as we do in Sec. IV. 

Before showing the details of this calculation we would like to comment on
the relation between the microscopic interaction \pref{Hamiltonian} and the
approximated one given in Eq. \pref{int}. If one restrict in the
interacting part of Eq. \pref{Hamiltonian} the sum over nearest-neighbors
sites one can easily show that $H_{int}$ can be rewritten as:
\be
\lb{eqint}
H_{int}=-V_0\sum_{\a,k,k',q \atop{\s\s'}} w_\a(\bk)w_\a(\bk')
c_{\bk\sigma}^\dagger c_{\bk+\bq\sigma'} 
c_{\bk'+\bq\sigma'}^\dagger c_{\bk'\sigma},
\ee
where the factor $(1/4)\sum_\delta e^{i(\bk-\bk')}$ coming from the sum
over nearest-neighboring sites $\delta$ has been decomposed in the
two-dimensional basis of wave functions $w_\a (\mathbf{k})$, which includes
the sum of contributions from several channels displaying different
symmetries with respect to the discrete rotation group for the square
lattice. One can see that even selecting only the $d$-wave channel and the
contribution at $\bq=\bQ$ Eq. \pref{eqint} does not contain only the
coupling in the particle-hole channel of Eq. \pref{int}, because the spin
structure in Eq. \pref{eqint} and Eq. \pref{int} are different. This
problem does not exist if the original microscopic model is given by a {\em
current-current} interaction, as the formation of a DDW state would
naturally require:
\be
\lb{intcc}
H_{int}=\frac{V_0}{2}\sum_{<i,j>\atop{\s\s'}}
c_{i\sigma}^\dagger c_{i\sigma'} 
c_{j\sigma'}^\dagger c_{j\sigma}=
-\frac{V_0}{2}\sum_{<i,j>\atop{\s\s'}}
c_{i\sigma}^\dagger c_{j\sigma} 
c_{j\sigma'}^\dagger c_{i\sigma'}.
\ee
Observe that: (i) Eq. \pref{intcc} is still gauge invariant, since the
Peierls transformation does not depend on the spin index; (ii) the
self-energy for the interaction \pref{intcc} is still given by
Eq. \pref{shf}, with $V(\bq)=2V_0$.  In the following we will never face
with the problem of solving explicitly Eq. \pref{sc1} for a given
microscopic interaction. However, it is worth noting that Eq. \pref{int}
can be directly derived by selecting a specific channel of a microscopic GI
model. Other examples can be also found in
Ref. \cite{Affleck:1988:PRB,Nersesyan:1989:JLTP,Schulz:1989:PRB,Varma:1997:PRB,Eremin:1998:JETPL,Cappelluti:1999:PRB}.

\subsection{The current-current correlation function evaluated with the
bare vertex $\gamma$}
\lb{sec:bare-vertex}

To evaluate the bubbles \pref{pmf} it is useful to introduce the
spectral representation for the Green's function \pref{Green-DDW}:
\be
\lb{spectral}
G(\bk, i \o_n)=\int_{-\infty}^{\infty} dz \frac{A(\bk,z)}{i\o_n-z}
\ee
with the spectral function
\be
\lb{addw}
A(\bk,z)=\frac{E_\bk+\e_\bk\s_3-D_\bk\s_2}{2E_\bk}\d(z-\mu-E_\bk)+
\frac{E_\bk-\e_\bk\s_3+D_\bk\s_2}{2E_\bk}\d(z-\mu+E_\bk).
\ee
The correlation functions
\pref{pmf} can then be written as
\be
\lb{pmf-DDW}
\Pi_{\m\n}^{(\gamma)}(\bq,i\O_m)=-\frac{2}{N}\sum^{RBZ}_\bk\int dz_1dz_2
Tr[A(\bk_+, z_1)\g_\m(\bk_+,\bk_-)
A(\bk_-,z_2)\g_\n(\bk_-,\bk_+)]\frac{f(z_1)-f(z_2)}{z_1-z_2-i\O_m}
\ee
which gives, according to \pref{addw}, the following
current-current correlation function:

\bea
\Pi_{ii}^{(\gamma)}(\bq,i\O_m)=
-\frac{1}{N}\sum^{RBZ}_\bk (v^F_{\bk_i})^2
\left[\frac{f(-\mu+E_+)-f(-\mu+E_-)}{E_+-E_- -i\O_m} +
\frac{f(-\mu-E_+)-f(-\mu-E_-)}{E_+-E_-+i\O_m}\right]
\left[1+\frac{\e_+\e_--D_+D_-}{E_+E_-} \right]\nn\\
\lb{pi-ii}
+(v^F_{\bk_i})^2
\left[\frac{f(-\mu+E_+)-f(-\mu-E_-)}{E_++E_--i\O_m} -
\frac{f(-\mu-E_+)-f(-\mu+E_-)}{E_++E_-+i\O_m}\right]
\left[1-\frac{\e_+\e_--D_+D_-}{E_+E_-} \right],\nn\\
\eea
where $D_\pm=D_{\bk\pm\bq/2}$, $\e_\pm=\e_{\bk\pm\bq/2}$,
$E_\pm=E_{\bk\pm\bq/2}$.

The issue then arises of the relation between the 
approximation \pref{pmf-DDW} for the correlation function and the 
sum rule \pref{deftmf}. Let us consider
again the GI relation \pref{GI}. When the static limit of the
current-current bubble is considered (corresponding in Matsubara
formalism to put $q=0$ in the expression \pref{pi-ii}), we find that
$\Pi^{(\g)}({\bf 0},0)$ is real and given by
\be
\lb{pi}
\Pi^{(\g)}_{ii}(\mathbf{0},0)=-\frac{2}{N}\sum^{RBZ}_\bk
\frac{(v^F_i)^2
D^2}{E^3}[f(\xi_+)- f(\xi_-)]+ \frac{(v^F_i)^2
\e^2}{E^2}[f'(\xi_+)+ f'(\xi_-)].
\ee
The usual procedure used to demonstrate that $\Pi({\bf 0},0)$ 
cancels out $<\tau>$
given by Eq.~\pref{deftmf} consists in integrating by parts the term in
$\Pi({\bf 0},0)$ which contains the derivative of the Fermi distribution
$f'(\xi_\pm)$ \cite{Scalapino:1993:PRB}. In order to do that
one would need a term like $df(\xi_\pm)/d\bk_i=\pm f'(\xi_\pm) (\e
v_{\bk_i}^F-D v_{\bk_i}^D)/E$, which in addition to the Fermi velocity,
$v^F_\bk$, contains also the DDW gap velocity, $v^D_{\bk_i}=-\pd
D_\bk/\pd\bk_i$. However, as one can easily see, the second term of
Eq.~\pref{pi} does not contain any contribution proportional to $v^D_i$, so
that the gauge-invariant relation $\Pi(\bq\ra 0,\o=0)=<\t>$ cannot be
satisfied with the bubble $\Pi^{(\g)}$.

According to the discussion of the previous section, the GI can only be
restored when the vertex corrections are included in the correlation
functions. Observe that the static-limit result \pref{gsol} reads in the
case of DDW state:
\be
\lb{gstatic}
\G_i(k,k)=\g_i(\bk, \bk)+v^D_{{\bf k}_i}\s_2=
v^F_{{\bf k}_i}\s_3+v^D_{{\bf k}_i}\s_2\equiv V_i({\bk }),
\ee
where $V_i(\bk)$ is the generalized velocity representing the $q=0$ limit
of the vertex function.  It is important to stress that in deriving
Eq. \pref{gsol} for $\G_i(k,k)$ it was crucial to keep the translation
invariant form $V(\bp-\bk)$ of the potential until the end. This point has
been often overlooked in the literature, at least while discussing
the corresponding problem for the superconducting case \cite{Schrieffer}.
If one used in Eq.~\pref{ver} defining the vertex function the
approximated form $V(\bk-\bp)\approx V_0 w_d(\bk)w_d(\bp)$, which is
appropriate for selecting  only the $d$-wave channel in the self-energy
\pref{shf}, the result \pref{gstatic} could not be obtained.
Indeed, a single-channel separable potential makes the
interaction term of the Hamiltonian {\em not} gauge invariant, and
then it would contribute to both the current operator and
diamagnetic term, as we will discuss in the next Section within the context
of the reduced model.  However, once that the result \pref{gsol}
has been established, and all the intermediate steps have been
performed in respecting GI requirements, we can
definitely select from the self-energy \pref{shf} only the
$d$-wave channel.  As a consequence, if $\Si(\bp)$ is approximated
as in Eq.~\pref{self-DDW}, the result \pref{gstatic} follows.

The Eq.~\pref{gstatic} can also be obtained from the generalized
WI \pref{wi}. Indeed, at small $\bq$ the difference
$G^{-1}(p)-G^{-1}(p+q)$, where $G$ is defined in
Eq.~\pref{Green-DDW}, is given by
\be 
\lb{wismallq}
q_\mu\G_\mu=G^{-1}(k)-G^{-1}(k+q)=-i\O_m
\s_0+v_\bk^F\cdot\bq\s_3+v_\bk^D\s_2 
\ee 
Since, according to the definitions \pref{vbare}, the bare vertex satisfies
$q_\mu\g_\mu=-i\O_m\s_0+v_\bk^F\cdot \bq \s_3$, for $\O_m=0$ we can find
again that the static vertex \pref{gstatic} satisfies the WI \pref{wi}.
Note that from the WI \pref{wismallq} one can be tempted to generalize the
result \pref{gstatic} for all $\O_m,\bq\approx 0$: however, one cannot
exclude that an additional term with zero space-time divergence can be
added to the solution \pref{gstatic}, still satisfying Eq.~\pref{wismallq}.

According to the discussion of Sec. II-D, one can try to use the
result \pref{gstatic} by evaluating the optical conductivity with
the symmetric bubble \pref{pisim}. In the specific case of the DDW
order, this would correspond to evaluating the following
current-current correlation functions: 
\be 
\lb{pigi-DDW}
\Pi^{DDW}_{ij}(\bq,i\O_m)=
-2\frac{T}{N}\sum^{RBZ}_{\bk,i\o_n}Tr[G(\bk_-,i\o_n+i\O_m)
V_i(\bk)G(\bk_+,i\o_n)V_j(\bk)]. 
\ee 
As it was explained in Sec. II D the ansatz \pref{pigi-DDW} guarantees the
correctness of the dc conductivity, and in general can be used to study the
low-frequency conductivity. Nevertheless, one can check that the bubble
$\Pi_{ii}^{DDW}(0)$ is not compatible with the diamagnetic tensor
\pref{deftmf}, violating again the GI condition \pref{GI} checked above for
$\Pi_{ii}^{(\gamma)}$.  The origin of this violation is obvious,
viz. instead of the asymmetric bubble \pref{pigi} with one full and one
bare vertex that would maintain the GI condition \pref{GI}, we used the
symmetric correlation function \pref{pisim}. Thus the issue arises whether
the diamagnetic tensor $\langle \tau \rangle$ can also be modified to
become compatible with the bubble \pref{pigi-DDW}.  As we shall see in the
next section, the diamagnetic tensor and sum rule corresponding to the
approximate bubble \pref{pigi-DDW} can be obtained without further
assumptions by analyzing the properties of the reduced Hamiltonian
\pref{DDW-Hamiltonian}.

\section{The reduced model}
\lb{sec:reduced}

An approach often proposed in the literature to deal with the DDW
state is that to consider directly the mean-field Hamiltonian
\pref{DDW-Hamiltonian} as the starting point
\cite{Benfatto:2003,Wang:2001:PRL,Zhu:2001:PRL, Kee:2002:PRB,
Yang:2002:PRB, Chakravarty:2002:PRL,Sharapov:2003:PRB,aristov}. The idea is
that at low energy the reduced model \pref{DDW-Hamiltonian} captures the
important physics of the system, so that one can consider it as a starting
microscopic Hamiltonian, describing non-interacting quasiparticles. In this
case the Green's function \pref{Green-DDW} does not provide any more an
approximation, but it is the correct one for the solvable, quadratic model
\pref{DDW-Hamiltonian}.  Since this Hamiltonian describes non-interacting
quasiparticles, it can be solved exactly and the corresponding conductivity
is given by the bare bubble.  This point of view was taken in
\cite{Benfatto:2003} where an unusual form of the optical-conductivity sum
rule was obtained.  One can notice that any distinction in the Hamiltonian
\pref{DDW-Hamiltonian} in the total energy between a kinetic and a
potential part, as can be done for the Hamiltonian \pref{Hamiltonian},
becomes somehow ambiguous, so that the result of \cite{Benfatto:2003} is
not surprising.

In what follows we compare this picture with the traditional one, and show
that since the dc conductivity calculated in both approaches appears to be
the same, one can also estimate the low-energy sum rule of the microscopic
model \pref{Hamiltonian} by considering the one realized in the reduced
model \pref{DDW-Hamiltonian}.

\subsection{The diamagnetic tensor, current operator and the sum rule
for the reduced model}

Let us now consider the Hamiltonian \pref{DDW-Hamiltonian} as the starting
microscopic model and analyze how all the considerations made in Sec. II
can be applied in this case.  Since the Hamiltonian \pref{DDW-Hamiltonian}
describes non-interacting quasiparticles, it is straightforward to
calculate the current-current correlation function and the electrical
conductivity, because in the absence of an interaction term the
Eq.~\pref{ver} for the vertex has a trivial solution $\Gamma_\mu (p_+, p_-)
= \tilde{\gamma}_\mu(p_+,p_-)$, where $\tilde{\gamma}_\mu(p_+,p_-)$ is the
bare vertex for the model DDW Hamiltonian \pref{DDW-Hamiltonian}.
Nevertheless, one should be careful and take into account that this vertex
is different from the bare vertex \pref{vbare} for the Hamiltonian
\pref{Hamiltonian}.  This can be understood by deriving the particle
current operator compatible with the conservation law (\ref{e-conserv}) and
with the equations of motion for the operators $c$ and $c^\dagger$,
\cite{Yang:2002:PRB,Kee:2002:PRB,Sharapov:2003:PRB,aristov}
\begin{equation}
\lb{electric.current.DDW}
\mathbf{j} (\mathbf{q},t)=\frac{1}{N}\sum_{\mathbf{k}, \sigma}
\left[ v_\bk^F c^{\dagger}_{\mathbf{k}- \mathbf{q}/2\s}
c_{\mathbf{k}+\mathbf{q}/2\s}
-iv_\bk^D c^{\dagger}_{\mathbf{k}- \mathbf{q}/2\s}
c_{\mathbf{k} + \mathbf{Q}+ \mathbf{q}/2\s} \right],
\end{equation}
The first term of the previous expression relates as usual the particle
current to the band velocity $v_\bk^F$. The second term, which only appears
for non-vanishing $D_0$, takes into account the contribution of the orbital
currents to the electrical conductivity, arising when the DDW order is
established. Observe that in the reduced model \pref{DDW-Hamiltonian} the
term proportional to $D_0$ appears as an additional, temperature dependent
band, which couples $\bk$ and $\bk+\bQ$ electrons, and as a consequence a
corresponding term appears in the definition of the current.  By rewriting
the electric current operator \pref{electric.current.DDW} using the spinors
\pref{spinors}, one has
\begin{equation}
\lb{electric.current.DDW-spinor}
\mathbf{j}_i (\mathbf{0},t)
 = \frac{1}{N}\sum_{\bk \sigma}^{\mathrm{RBZ}}
\chi_{\mathbf{k} \s}^{\dagger}  V_i(\mathbf{k})
 \chi_{\mathbf{k} \s}
\end{equation}
and, accordingly, the bare vertex reads \cite{Sharapov:2003:PRB}:
\be
\lb{vbare-DDW}
\tilde{\gamma}_\mu(\bk-\bq/2,\bk+\bq/2)=( V_i({\bk}), \s_0), \quad
\bq\to 0,
\ee
where $V_i(\bk)$ is the generalized velocity defined in Eq.~\pref{gstatic}.
Substituting the bare vertex \pref{vbare-DDW} and the Green's function
\pref{Green-DDW} in the Ward identity \pref{WI-dif} one can easily see that
it is satisfied.  Moreover, since for non-interacting quasiparticles the
full and bare vertex functions coincide, the correlation function
$\Pi^{(\tilde \g)}$ of Eq.~\pref{pmf}, evaluated with the bare vertex
$\tilde \g$ of Eq.~\pref{vbare-DDW}, has two properties: (i) it is the {\em
exact} one for the quadratic model \pref{DDW-Hamiltonian}; (ii) it coincides
with $\Pi^{DDW}_{ij}$ in Eq.  \pref{pigi-DDW}, which is an {\em
approximation} for the full model \pref{Hamiltonian}. As a consequence, the
sum rule corresponding to the bubble $\Pi^{DDW}_{ij}$ can be obtained by
the knowledge of the stress tensor for the reduced system.  Observe that
the current operator \pref{electric.current.DDW} is also obtained when the
Peierls substitution is performed directly in the reduced model
\pref{DDW-Hamiltonian}. As we discussed in
Sec.~\ref{sec:general-sigma-sum}, after the Peierls substitution both the
current operator and the diamagnetic tensor can be derived from $H(A)$,
according to Eq.~\pref{j-tau.def}. As a consequence, in the reduced model
not only the current operator but also the diamagnetic tensor $\tau_{ii}$
is modified, containing an extra term for $D_0\neq 0$ \cite{Benfatto:2003},

\be
\lb{new.tensor}
\langle \t_{ii}\rangle=-\frac{1}{2N}\sum_{\bk\s} \left[\e_\bk \langle
c_{\bk\s}^\dagger c_{\bk\s} \rangle+ i D_\bk \langle
c_{\bk\s}^\dagger c_{\bk+\bQ\s}\rangle \right].
\ee
When the operator averages are evaluated, or analogously
Eq.~\pref{Enz.identity} is used, one finds that the sum rule
for the reduced model is:
\begin{equation}
\lb{sum.T>Tc}
\frac{W^{DDW}(D,T)}{(\pi e^2a^2/V)}=-
\frac{1}{N}\sum_{\bk}^{RBZ}
E_\mathbf{k}
[f (\xi_{+,\mathbf{k}}) - f(\xi_{-,\mathbf{k}})],
\end{equation}
where $E_\mathbf{k}$ and
$\xi_{\pm,\mathbf{k}}$ were already defined after Eq.~\pref{sc2}.
Eq.~(\ref{sum.T>Tc}) was derived using the fact that
$\partial_{x,y} v^F_\bk= 2ta^2 \cos k_{x,y} a$ (and
$\partial_{x,y} v^D_\bk= \pm (D_0/2) a^2 \cos k_{x,y}a$), and it reduces to
Eq.~(\ref{weight.band}) for $D_0=0$.

Once more, the result \pref{new.tensor} is consistent with the GI
for the reduced model. Indeed, if the bubbles $\Pi^{DDW}(\bq,i\O_m)$ 
\pref{pigi-DDW} are evaluated in the static limit, instead of the result 
\pref{pi} for $\Pi^{(\g)}$  one has: 
\be 
\lb{pigi0}
\Pi_{ii}^{DDW}({\bf 0},0)=-\frac{2}{N}\sum_\bk^{RBZ} \frac{(v_i^F
D+v_i^D\e)^2}{E^3}[f(\xi_+)- f(\xi_-)]+ \frac{(v_i^F \e-v_i^D
D)^2}{E^2}[f'(\xi_+)+ f'(\xi_-)] 
\ee 
If now one integrates by parts the second term of Eq.~\pref{pigi0} one
finds that: 
\bea
-\frac{2}{N}\sum_\bk^{RBZ}\frac{(v_i^F \e-v_i^D
D)^2}{E^2}[f'(\xi_+)+
f'(\xi_-)]=\frac{2}{N}\sum_\bk^{RBZ}[f(\xi_+)-f(\xi_-)]\frac{\pd}{\pd\bk}
\frac{(v_i^F \e-v_i^D D)}{E}=\nn\\
\frac{2}{N}\sum_\bk^{RBZ} \frac{(v_i^F D+v_i^D\e)^2}{E^3}[f(\xi_+)-f(\xi_-)]
-\frac{1}{N}\sum_{\bk}^{RBZ}E
[f (\xi_{+}) - f(\xi_-)],\nn
\eea
and as a consequence the GI relation $\Pi_{ii}^{DDW}({\bf 0},0)=\langle 
\t_{ii} \rangle$ with $\langle \t_{ii} \rangle$ given by Eq.~\pref{new.tensor}
is satisfied, as expected when the exact vertex $\G$ in included in the bubble.

A comment is in order now about a third possible approach proposed
in the literature \cite{Chakravarty:2002:PRL} for the analysis of
the reduced model \pref{DDW-Hamiltonian}.
By rewriting the
quadratic Hamiltonian \pref{DDW-Hamiltonian} as $
H=\sum_{\bk,\s}^{RBZ}\chi_{\bk\s}^+\hat H_\bk \chi_{\bk\s} $ the
matrix $\hat H_\bk$ can be diagonalized by means of an unitary
transformation $U_\bk$, $\hat H=U_\bk\Lambda_\bk U_\bk^+$ where
$\Lambda_\bk=diag(\xi_+,\xi_-)$. According to our definition
\pref{electric.current.DDW}, the current is derived from $\hat
H_\bk({\bf A})$, so that it corresponds to
$j_{DDW}=(1/N)\sum_{\bk\s}^{RBZ} \chi_{\bk\s}^+(\pd_\bk\hat H_\bk)
\chi_{\bk\s}= (1/N)\sum_{\bk\s}^{RBZ} \chi_{\bk\s}^+\pd_\bk(U_\bk\Lambda_\bk
U_\bk^+) \chi_{\bk\s}$ (see also \cite{aristov}). Let us introduce the
spinors $\psi_{\bk\s}=U_\bk^+\chi_{\bk\s}$ which diagonalize the
Hamiltonian matrix $\hat H$,
$H=\sum_{\bk\s}^{RBZ}\psi_{\bk\s}\Lambda_\bk\psi_{\bk\s}$. Then, by
making the assumption that the gauge field couples by Peierls ansatz not to
$\chi_{\bk\s}$ but to the new quasiparticle
operators $\psi_{\bk\s}$, one would calculate the current starting from
$\Lambda_\bk({\bf A})$, so that the current, the diamagnetic term and
the static limit of the current-current bubble would be defined as
\cite{Chakravarty:2002:PRL}:
\bea
j_{QP}=\frac{1}{N}
\sum_{\bk\s}^{RBZ}\psi_{\bk\s}\pd_\bk \Lambda_\bk \psi_{\bk\s}=
\frac{1}{N} \sum_{\bk\s}^{RBZ}\chi_{\bk\s}^+U_\bk(\pd_\bk\Lambda)
U_\bk^+ \chi_{\bk\s},\nn\\
<\tau_{ii,QP}>=\frac{1}{N}\sum_{\bk\s}^{RBZ}
\left[\frac{\pd^2 \xi_+}{\pd k_i^2}
f(\xi_+) +\frac{\pd^2 \xi_-}{\pd k_i^2} f(\xi_-)\right],\nn\\
\Pi_{ii,QP}({\bf 0},0)=-\frac{1}{N}\sum_{\bk\s}^{RBZ}
\left[\left(\frac{\pd \xi_+}{\pd k_i }\right)^2 f'(\xi_+)+
\left(\frac{\pd \xi_-}{\pd k_i }\right)^2 f'(\xi_-)\right].\nn
\eea
Observe that this approximation is still GI in the sense that it is easy to
see that $\tau_{ii,QP}$ and $\Pi_{ii,QP}(0)$ defined above satisfy the
condition \pref{GI}. However, this approximation has no relation with the
microscopic starting model, in the way we explained in
Sec.~\ref{sec:sum-rule-Hubbard}.  For this reason, we do not comment
further on this approach, and we analyze instead the result obtained with
the current operator \pref{electric.current.DDW} and the bubbles
\pref{pigi-DDW}, whose correspondence with the microscopic model we
established above.

\subsection{Temperature dependence of the spectral weight}

Once that we clarified the different approximations used in deriving the
two sum rules \pref{deftmf} and \pref{sum.T>Tc}, let us discuss the
outcomes of these two approaches as far as the temperature dependence of
the spectral weight is concerned.  Two observations should be kept in mind:
(i) the overall variations of the spectral weight in the DDW state are not
expected in general to be large if quite small gap values $D(0)/t\ll 1$ are
considered; (ii) in Eqs.~\pref{deftmf} and \pref{sum.T>Tc} the temperature
variation of both the gap and the chemical potential $\mu$ contribute to
the shape of $W(T)$. In the case of free electrons, the variation of
$\mu(T)$ is almost negligible compared to the temperature variation of the
occupation number, given by the Fermi function. Indeed, even considering
the temperature variation of $\mu(T)$ the result \pref{weight.normal} is
only modified by terms of order $T^4$.  In the case of
Eqs.~\pref{deftmf} and \pref{sum.T>Tc} also the band structure is varying in
temperature, and it is important to keep track of this by solving at each
temperature the self-consistent equation for the chemical potential. Here,
instead of solving explicitly Eq.~\pref{sc1}, we adopt a general mean-field
temperature dependence for $D_0(T)=D(0)g(T/T_{DDW})$, with
$g(x)=(1-x^4/3)\sqrt{1-x^4}$, as shown in the left panel of Fig.~1.  In the
right panel of Fig. 1 we also present the temperature dependence of the
chemical potential in the DDW state: as one can see, below $T_{DDW}$ there
is an inversion of tendency of $\mu(T)$ due to the opening of the gap. The
temperature dependence of the spectral weight in the DDW state according to
Eq. \pref{deftmf} ($W(D,T)$) and \pref{sum.T>Tc} ($W^{DDW}(D,T)$) is
reported in Fig. 2, where also the tight-binding spectral weight
\pref{weight.normal} ($W(T)$) is shown for comparison. Here we used a small gap
value, $D(0)=2.5 T_{DDW}=0.3 t$, and doping $\d=0.1$.  The influence of the
chemical-potential variation are evident comparing the right panel of
Fig. 2, where $W(D,T)$ and $W^{DDW}(D,T)$ are evaluated keeping $\mu$
constant (at the value it has in the normal state), and the left panel,
where the density is constant. In addition, we see that for this value of
$D(0)$ the overall spectral-weight variations are small in the DDW state.
However, it is found that the definition \pref{deftmf} leads to a smooth
decrease of the spectral weight below $T_{DDW}$, in analogy with the
results for a SC transition, while the definition \pref{sum.T>Tc} gives an
increase.  Such variations are quantitatively (but not qualitatively)
modified if the temperature variations of the chemical potential are not
properly taken into account, see right panel of Fig. 2. Observe that the 
relative variations of $W(T)$ between $T=0.16 t$ and $T=0$ are never larger 
than $\sim 1.2 \% $, and cannot be appreciated on the scale of the figure
reported in Ref. \cite{aristov}.

\begin{figure}[htb]
\centering{
\includegraphics[width=8.cm,angle=-90]{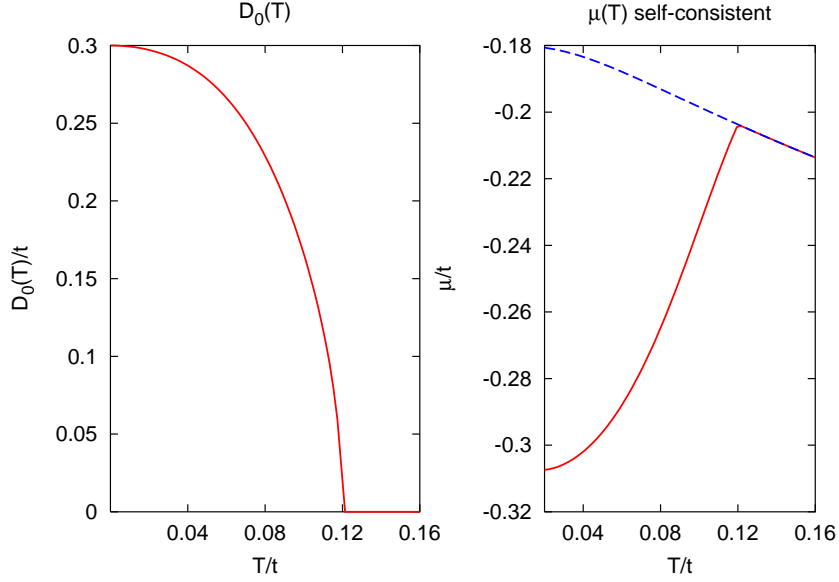}
} \caption{(Color online) Left panel: Temperature dependence of the DDW gap
according to the function $g(x)$ defined in the text. Right panel:
$\mu(T)$ in the normal state (dashed line) and in the DDW state
(solid line), obtained solving the self-consistency equation \pref{sc2} for
the particle number (with $D_0=0$ for the normal state).}
\label{fig:1}
\end{figure}

Even though a detailed description of cuprates is not the main aim of our
paper, we find nevertheless useful to compare our results for a choice of
parameters appropriated for HTSC.  Since on this respect different
attitudes are present in the literature, we briefly recall here the phase
diagram analyzed in Ref.~\cite{Benfatto:2000:EPJ} within the more general
attitude of investigating the consequences of describing the pseudogap
state with a $k$-space modulated charge density wave. In
Ref. \cite{Benfatto:2000:EPJ} it was shown that one outcome of this
description is the possibility to interpret the leading-edge shift observed
in photoemission experiments as due to a particle-hole gap. In particular,
for a band dispersion with a next-nearest neighbors hopping term $t'=0$ the
hole-pockets Fermi surface formed by doping the DDW system with respect to
half-filling is a simplification intended to reproduce the arcs of Fermi
surface observed experimentally. A simple calculation shows that in such a
case the gap measured by ARPES at the M points corresponds approximately to
$D_0-|\mu|$. As a consequence, $D_0(0)$ and $T_{DDW}$ do not correspond
directly to the maximum gap value and the $T^*$ temperature measured by
ARPES, but both are quite larger, as shown in Ref.~\cite{Benfatto:2000:EPJ}
where the values of the DDW gap and of the temperature $T_{DDW}$ were
chosen to properly reproduce the phase diagram of Bi2212 compounds.

\begin{figure}[htb]
\centering{
\includegraphics[width=6.cm,angle=-90]{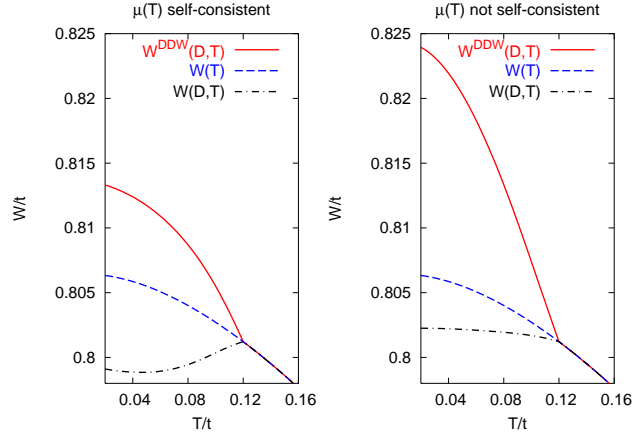}}
\caption{(Color online) Spectral weight $W^{DDW}(D,T)$, $W(D,T)$, and
$W(T)$, according to Eqs. \pref{sum.T>Tc}, \pref{deftmf}, and
\pref{weight.normal}, respectively, in units of $e^2\pi a^2/V$. We used
here $D(0)=2.5 T_{DDW}= 0.3 t$ and $\d=0.1$. Left panel: results obtained
using the chemical potential obtained solving the self-consistency equation
\pref{sc2} for the particle number with $D_0\neq 0$ (for $W^{DDW}(D,T)$ and
$W(D,T)$) and $D_0=0$ (for $W(T)$) respectively, see Fig. 1. Right panel:
evaluation of $W^{DDW}(D,T)$ and $W(D,T)$ using the chemical potential of
the normal state.}
\label{fig:2}
\end{figure}

In agreement with Ref.~\cite{Benfatto:2003} we keep here this
attitude and use a doping and temperature dependent DDW gap
$D_0(T,\delta)=cT_{DDW}(\delta)g(T/T_{DDW})$, where
$T_{DDW}(\delta)=0.16t[1-(\delta/\delta_0)^4]$ vanishes at the
critical doping $\d_0=0.2$ for DDW formation, and $c=7$ is a fitting
parameter \cite{Benfatto:2000:EPJ}. Since the resulting temperature
dependence of the sum rule \pref{sum.T>Tc} was already shown in
Ref. \cite{Benfatto:2003}, we just report here for comparison the
behavior of the two sum rules $W(D,T)$ in Eq.~\pref{deftmf} and
$W^{DDW}(D,T)$ in Eq.~\pref{sum.T>Tc}
for this choice of parameters at $\d=0.16$. As it can be seen in
Fig. 3, the relative variation of the spectral weight below
$T_{DDW}$ is made now more pronounced, enhancing the differences
between the two possible approaches followed in deriving the sum
rule. It is then clear that the standard sum-rule derivation
leading to $W(D,T)$ in Eq.~\pref{deftmf} cannot be consistent with
the experiments, since no decrease of the spectral weight has been
observed in the pseudogap phase of cuprates. The result $W^{DDW}(D,T)$ in
Eq.~\pref{sum.T>Tc}  is instead resembling more closely the
experimental findings, in particular if we consider that at this
doping level the room temperature below which the data in Ref.
\cite{Molegraaf:2002:Science,Santander:2002:PRL,Santander:2004}
are reported corresponds to $T/t\sim 0.1$, so that the overall
measured temperature dependence of $W(T)$ would correspond in our
picture to the DDW result \pref{sum.T>Tc}. Indeed, as we show in
the right panel of Fig. 3, the $W^{DDW}(D,T)$ evaluated according to
Eq.~\pref{sum.T>Tc} still displays a $T^2$ temperature dependence,
but with a larger slope, as observed experimentally. This approach
would allow one to understand why the spectral-weight increase
looks like a ``standard'' free tight-binding model, but with a
much larger slope.  However, as we shall see in the
next subsection, the comparison with the experiments is made much
more involved when the optical conductivity corresponding to the
sum rule \pref{sum.T>Tc} is evaluated. Finally, one can in
principle extend this analysis to the case where also SC is added,
but since also the experimental situation is not clear on this
respect we refer to Ref.~\cite{Benfatto:2003} for a discussion
about the SC state.

\begin{figure}[htb]
\centering{
\includegraphics[width=6.cm,angle=-90]{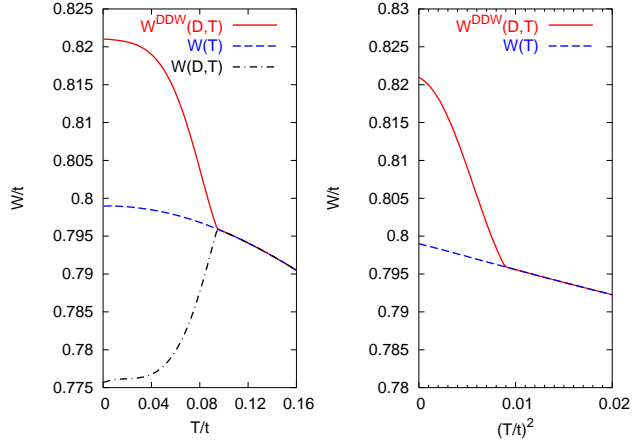}}
\caption{(Color online) Left panel: Spectral weight $W^{DDW}(D,T)$, $W(T)$,
and $W(D,T)$ in units of $e^2\pi a^2/V$ for a choice of parameter values
appropriate for cuprates (see the discussion in the text). Here we show the
results for $\delta=0.16$, corresponding approximately to optimal doping,
and we calculate the chemical potential self-consistently. Right panel:
spectral weight plotted as a function of $(T/t)^2$: one can see that a
$T^2$ temperature dependence is still recovered for $W^{DDW}(D,T)$ in a
wide range of temperatures.}
\label{fig:3}
\end{figure}

\subsection{The role of a next-nearest neighbors hopping term}
Up to know we did not consider the possibility of a next-neighbors hopping
term $t'$ in the bare band dispersion $\e_\bk$. Indeed, from one side we
wanted to simplify the notation while discussing the issue of the relation
between gauge invariance and sum rule, and from the other side we believe
that even when comparing with cuprates the case $t'=0$ is enough to reproduce
phenomenologically the arc of Fermi surface observed in the pseudogap phase
(see discussion above). However, for the sake of completeness, we report
here briefly the modifications induced in the sum rule when a $t'$ term is
included in the band dispersion, so that
\bea
\e_\bk=s_\bk+p_\bk,\nn\\
s_\bk=-2t (\cos k_x a +\cos k_y a)\nn\\
\lb{etp}
p_\bk= 4t' \cos k_x a \cos k_y a.
\eea
In the DDW state the perfect nesting condition is lost due to the $t'$
term, so that $\e_\bk+\e_{\bk+\bQ}=2p_\bk, \e_\bk-\e_{\bk+\bQ}=2s_\bk$
and the two quasiparticle branches in the DDW state become
$\xi_{\pm,\bk}=p_\bk-\mu\pm E_\bk$, where $E_\bk=\sqrt{s_\bk^2+D_\bk^2}$. 
As a consequence, given the relation
\pref{common.rule} between the sum rule and the diamagnetic tensor, and the
definitions \pref{deft} and \pref{new.tensor} of the diamagnetic tensor for
the original and the reduced model respectively, it is easy to see that
Eqs. \pref{weight.normal}, \pref{deftmf} and \pref{sum.T>Tc} get modified
as:
\bea
\lb{wntp}
\frac{W(T)}{(\pi e^2 a^2/V)}&=&-\frac{1}{N}\sum_\bk (\e+p) f(\xi)\\
\lb{tmftp}
\frac{W(D,T)}{(\pi e^2 a^2/V)}&=&-\frac{1}{N}\sum^{RBZ}_\bk \left\{
\frac{s^2}{E}[f(\xi_+)-f(\xi_-)]+2p[f(\xi_+)+f(\xi_-)]\right\},\\
\lb{sumtp}
\frac{W^{DDW}(D,T)}{(\pi e^2a^2/V)}&=&-
\frac{1}{N}\sum_{\bk}^{RBZ} \left\{E [f (\xi_{+}) -  f(\xi_{-})]
+2p[f(\xi_+)+f(\xi_-)]\right\} ,
\eea
where the explicit dependence on $\bk$ is omitted. In Fig. 4 we
compare again the temperature dependence of the spectral weight in the
different cases, for  $t'=0.3t$, $\delta=0.1$, $T_{DDW}=0.12t$ and
$D(0)=4T_{DDW}$. Even though the introduction of the $t'$ term modifies the
temperature dependence of the chemical potential in the normal and DDW
state (due to the shift of the Van Hove singularity which is now below the
Fermi level at $\delta=0.1$), the general trend of Figs. 2-3 is
confirmed. Indeed, $W^{DDW}(D,T)$ is larger than $W(T)$ below $T_{DDW}$,
while $W(D,T)$ is smaller. In particular, it is worth noting that
apart from possible quantitative differences with respect to the case
$t'=0$, the exact form of the band dispersion is irrelevant as far as
the main issue discussed in the previous sections, i.e. the fact that
different approximations for the current-current correlation functions lead
to different results for the optical-conductivity sum rule. For this
reason, we do not discuss further in the following the role of a $t'$ term,
and remind the reader for example to Refs. \cite{aristov,carbotte}, 
where this issue is investigated in
more details.

\begin{figure}[htb]
\centering{
\includegraphics[width=6.cm,angle=-90]{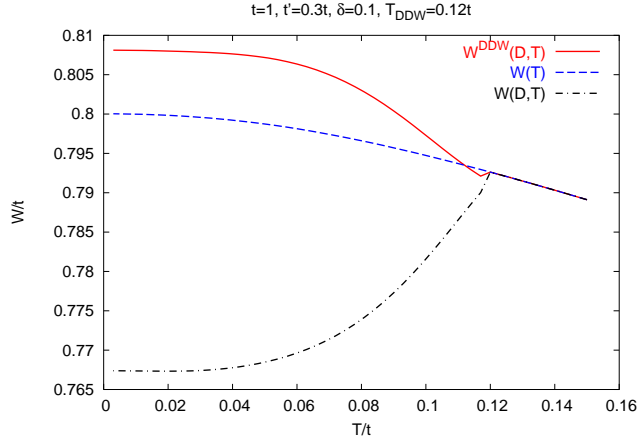}}
\caption{(Color online) Spectral weight in the presence of a $t'$ term in
  the band dispersion. Here we show $W^{DDW}(D,T)$, $W(D,T)$, and $W(T)$,
  according to Eqs. \pref{sumtp}, \pref{tmftp}, and \pref{wntp},
  respectively, in units of $e^2\pi a^2/V$. We used here $t'=0.3t$,
  $T_{DDW}=0.12t$, $D(0)=4 T_{DDW}$ and $\d=0.1$. The chemical potential is
  evaluated self-consistently at each temperature by solving Eq. \pref{sc2}
  in the presence of a $t'$ term in the band dispersion. Observe that near
  $T_{DDW}$ a small decrease of $W^{DDW}$ with respect to $W(T)$ is
  observed, due to the change of chemical potential near $T_{DDW}$.}
\label{fig:4}
\end{figure}

\subsection{The optical conductivity of the reduced model}
\lb{sec:conductivity-DDW}

As we discussed in the previous sections, one would expect that our
result \pref{sum.T>Tc} for $W^{DDW}(D,T)$ is only valid at low energy scales,
possibly below the plasma frequency, which can be thought as a general
cut-off for any tight-binding based description of the system. In principle
one could also expect that the low-energy theory \pref{DDW-Hamiltonian} is
valid at even lower energy scales, but since at the plasma energy one still
finds experimentally strong variation with respect to the naive estimate
\pref{weight.normal}, it is plausible that a quite larger cut-off holds
here for the tight-binding model itself. To analyze the dependence of the
result \pref{sum.T>Tc} on the cut-off frequency we need an explicit
calculation of the optical conductivity obtained with the bubble
\pref{pigi-DDW}.

By using the spectral representation of the Green's functions the
current-current correlation function $\Pi^{DDW}_{ii}$ \pref{pigi-DDW} can
be evaluated in analogy with $\Pi^{(\g)}_{ii}$ in Eq.~\pref{pmf-DDW}, with
the bare vertices $\g_i$ substituted by the full one $\G_i(k,k)$ of Eq.~
\pref{gstatic}. To take into account the effect of disorder we make the
simplest ansatz of substituting the delta functions associated to a
quasiparticle pole in the spectral representation \pref{addw} with a
Lorentzian of finite width $w$ ($w = 1/(2 \tau_{tr})$, where $\tau_{tr}$ is
the transport time):

\be
\lb{dis}
\d(z)\ra
M(z)=\frac{1}{\pi}\frac{w}{z^2+w^2}.
\ee
As a consequence, after
analytical continuation in Eq.~\pref{pmf-DDW}, we obtain:
\begin{equation}
\lb{sigma}
\begin{split}
\s^{DDW}(\o)= & \\  -\frac{2\pi e^2}{V} \sum^{RBZ}_\bk\int dz
\frac{f(z+\o)-f(z)}{\o} & \left\{\frac{(\e v^F-Dv^D)^2}{E^2}
[M(z+\o-\xi_+)M(z-\xi_+)+M(z+\o-\xi_-)M(z-\xi_-)]\right.\\
 + & \left.\frac{(\e v^D+Dv^F)^2}{E^2}
[M(z+\o-\xi_+)M(z-\xi_-)+M(z+\o-\xi_-)M(z-\xi_+)]\right\},
\end{split}
\end{equation}
where $v^F$ and $v^D$ refers to the component in a given x or y direction.

As already observed in Ref. \cite{Yang:2002:PRB}, and more recently in
Refs. \cite{aristov,carbotte}, the optical conductivity
is composed of two contributions, due to the splitting of the original
single band $\e_\bk$ in two new bands $\xi_\pm$ after the gap opening.  In
Eq. \pref{sigma} the first line describes intra-band excitations
(corresponding to the product of two $M$ functions evaluated at the same
quasiparticle branch), while the second line takes into account inter-band
processes. It is easy to see that this second contribution is only possible
when $\o>2|\mu|$ (at low temperatures). Indeed, when the system is doped
with respect to half-filling ($|\mu|\neq 0$) the smallest energy difference
between occupied and unoccupied states in different branches is equal to
$2|\mu|$, and it is realized at the points $(\pm \pi/2, \pm \pi/2)$ where
the energy $E_\bk$ vanishes and the two bands merge. The first contribution
has instead a Drude-like shape, as it is shown in Fig. 5, where we report
the optical conductivity at $T=0$ for a system with and without DDW
gap. Here we used the set of parameters discussed above for cuprates, at
$\d=0.13$. When compared with the free-electron conductivity at the same
temperature, one can see that the Drude peak is smaller in the DDW state,
because part of the spectral weight has been transferred to the inter-band
processes. To quantify this transfer of spectral weight we integrate
numerically the optical conductivity $\s^{DDW}(\o)$ \pref{sigma}, and its
analogous $\s(\o)$ at $D_0=0$, evaluating for a given cut-off frequency
$\o$ the quantity:
\be
\lb{no}
N^{(DDW)}(\o)=2\int_0^\o \s^{(DDW)}(\o')d\o',
\ee
which verifies $N^{DDW}(\o\ra \infty)=W^{DDW}(D,T=0)$ and $N(\o\ra \infty)
\ra W(T=0)$ for the DDW and normal state respectively.

\begin{figure}[htb]
\centering{
\includegraphics[width=8.cm,angle=-90]{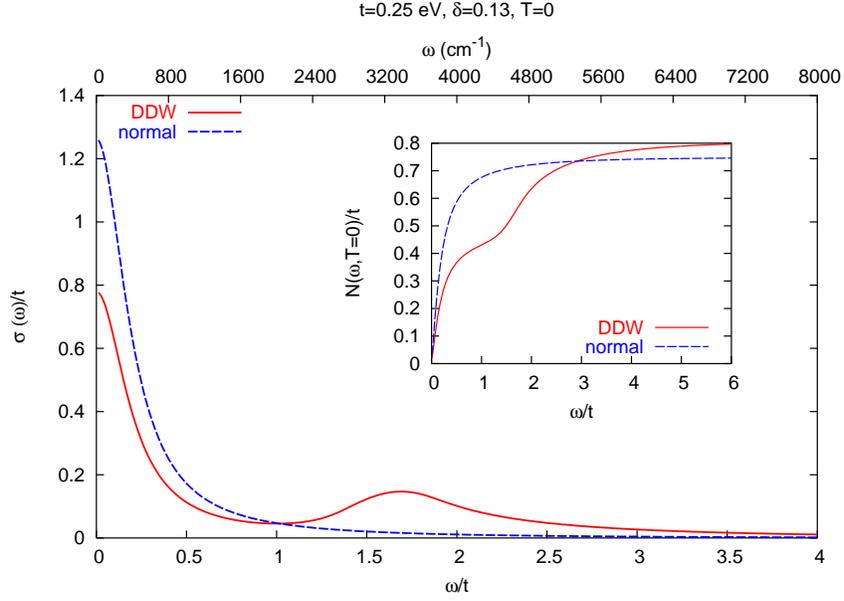}
} \caption{(Color online) Optical conductivity in units of $e^2\pi a^2/V$
at zero temperature for a free
tight-binding system and for the DDW state at $\delta=0.13$
($\mu=-0.69t$ at $T=0$ in the DDW state, $w=0.1t$ were
used). For convenience we also report the frequencies in
cm$^{-1}$, as it is customary in the experiments. Inset: frequency
variation of $N(\o)$ according to Eq.~\pref{no}. Observe that at
low cut-off energy the spectral weight in the DDW state is smaller
than in the normal state.} \label{fig:5}
\end{figure}

In the inset of Fig. 5 we show $N^{DDW}(\o)$ and $N(\o)$ at $T=0$
corresponding to the calculated optical conductivities. As we can see, at
low energy the formation of a DDW state leads to an overall decrease of
spectral weight, since intra-band processes are partly suppressed. However,
at higher energy inter-band excitations are allowed and the spectral weight
lost in the Drude peak is over-compensated, giving rise to an overall
increase of $W^{DDW}(D,T=0)$ in the DDW state compared to $W(T=0)$ in the
normal state. For a value of $t\sim 0.25$ eV one sees that in the case of
Fig. 5 the crossing of $N^{DDW}(\o)$ with respect to $N(\o)$ is already
satisfied at cut-off frequencies smaller than the plasma frequency
($\approx 4t$), even though $N^{DDW}$ saturates at higher frequencies.  Of
course it is evident that the determination of the exact frequency at which
the sum rule $W(T)$ or $W^{DDW}(T)$ are exhausted depends on the choice of
parameters.  For example, at smaller doping or smaller $D_0$ (which both
lead to a smaller value of $|\mu|$ in the DDW state, see Fig. 1) the
intra-band processes occur at lower energy, so that $N^{DDW}(\o)>N(\o)$
will be satisfied at lower cut-off energy $\o$. In Fig. 5 we report one of
the cases when the cut-off frequency is larger, because the chemical
potential shifts from $\mu=-0.28t$ in the normal state to $\mu=-0.69t$ in
the DDW state, due to the large value of the DDW parameter ($D_0=0.92t$),
pushing inter-band processes at relatively high energies. It is interesting
to observe that the optical conductivity in the DDW state reported in
Fig. 5, which was evaluated with the ansatz \pref{pigi-DDW} for the
current-current correlation function, has the same {\em qualitative}
behavior of the optical conductivity reported in Ref. \cite{carbotte},
where the bare bubble approximation \pref{pmf} was considered for the
correlation function. Indeed, in the bare bubble approximation the optical
conductivity has the same structure of Eq. \pref{sigma} (i.e. a Drude term
plus inter-band processes), but with $v_D=0$. However, the two approaches
lead to two {\em quantitatively} different temperature dependences of the
sum rule. Indeed in the absence of the $v_D$ term in Eq. \pref{sigma},
coming from the vertex corrections, the spectral weight lost in the Drude
term when the DDW state is fromed would not be compensated any more by the
inter-band processes, so that the total spectral weight
$N^{DDW}(\o\rightarrow \infty)$ in the DDW state would be always lower than
the spectral weight $N(\o\rightarrow\infty)$ in the normal state, as
observed in Ref. \cite{carbotte}.

\begin{figure}[htb]
\centering{ \includegraphics[width=8.cm,angle=-90]{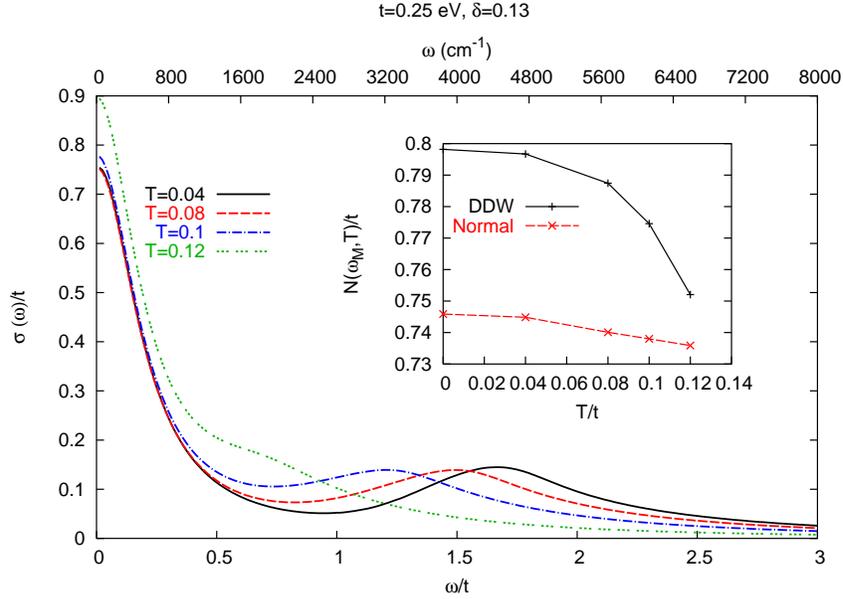}
}\caption{(Color online) Optical conductivity in units of $e^2\pi a^2/V$ at
various temperature for the DDW state at $\delta=0.13$, $T_{DDW}=0.13t$. In
the upper x-axis the frequencies are reported in cm$^{-1}$. Inset:
comparison between the temperature dependence of the integrated spectral
weight $N^{DDW}(\o_M)$ and $N(\o_M)$ for the DDW and normal state
respectively. The cut-off is $\o_M=6t$.}
\label{fig:6}
\end{figure}

In Fig.\ 6 we report the optical conductivity in the DDW state at several
temperature between $T=0$ and $T_{DDW}=0.13t$ at the doping
$\delta=0.13$. As one can see, when the temperature increases the
inter-band processes shift to lower frequency, due to the decrease of the
absolute value of the chemical potential (see right panel of Fig.\ 1). As a
consequence, the spectral weight is transferred again towards the Drude
peak, and the overall balance of spectral weight leads to a decrease of
$W^{DDW}(D,T)$. In the inset of Fig.\ 6 we show also the integrated
spectral weight $N^{DDW}(\o_M)$ and $N(\o_M)$ at the same temperatures of
the main panel, with a cut-off frequency $\omega_M=6t$. Even though this
estimate of the spectral weight is much less accurate than the direct
evaluation of Eq. (\ref{sum.T>Tc}), due to lowest numerical accuracy of
this procedure, we find the same behavior discussed in the previous
Sections while computing directly Eq. (\ref{sum.T>Tc}). Indeed, we can see
that the spectral weight increases in the DDW state with respect to the
normal state, even though the relative contribution of the Drude term is
lower in the DDW state than in the normal state.

A comment is in order now about the role of disorder. In the previous
sections we reported the numerical results obtained for clean systems, but
to amplify the differences between the conductivity of a non-interacting
system and of the DDW state we used in Figs. 5, 6 a quite large value of the
inverse scattering time $w=0.1t$, as appropriate for example to reproduce
qualitatively the large Drude peak observed in BSCCO samples at about 100K
\cite{Santander:2002:PRL,Santander:2004}. As a consequence, also the
self-consistency equation \pref{sc2} for the particle number and the sum
rules $W^{DDW}(D,T)$ and $W(T)$ should be evaluated in the presence of
disorder for a given doping. The main difference is only in the absolute
value of the spectral weight, while the relative difference between the
case with or without DDW is the same. In the appendix B we discuss the
modifications to the particle-number and spectral weight equations in the
presence of disorder, that we used in computing the optical conductivity in
Figs. 5, 6.

\section{Discussion}

In the present paper we analyzed a possible approach to determine a GI
approximation for the optical conductivity in a system which displays a
transition to a d-wave modulated CDW or flux phase. As we explained in
detail in Sec.~\ref{sec:sum-rule-Hubbard} the requirement of GI of a theory
fixes the relation \pref{GI} between the current-current correlation
function and the diamagnetic term. To understand better the expected
domain of applicability the sum rule $W^{DDW}(D,T)$ \pref{sum.T>Tc} let us
summarize the assumptions that led us to this rule. We have checked in
Sec.~\ref{sec:bare-vertex} (Eq.~\pref{pi-ii}) that the bubble $\Pi^{(\g)}$
of Eq.  \pref{pmf}, evaluated with the mean-field DDW Green's function
\pref{Green-DDW} and the bare vertex $\gamma$, does not satisfy the GI
condition \pref{GI} when the standard diamagnetic term \pref{deftmf} is
considered.  As discussed in Sec.~\ref{sec:WI}, this situation is quite
standard, and considering the WI \pref{gsol} this violation of the gauge
invariance can be attributed to the $\bk$-dependent character of the DDW
gap $D_\bk$, which makes necessary the use of the full vertex $\Gamma
(k,k)$ instead of the bare one $\gamma(\bk,\bk)$.

In general the vertex function is determined by solving the
integral equation \pref{ver}, but in the static limit it reduces
to the expression \pref{gsol}.  Since we do not know an analytical
solution at finite frequencies and momenta of the vertex
$\Gamma(k_+,k_-)$, corresponding to the microscopic many-body
Hamiltonian \pref{Hamiltonian}, we can try to use our knowledge of
its static limit to give a better approximation than
Eq.~\pref{pmf} for the current-current correlation function. More
precisely, we showed in Sec.~\ref{sec:sigma-sym} that the dc
conductivity derived from the symmetric bubble $\Pi_{ij}^{sym}$
\pref{pisim}, where two full vertex functions in the static limit
appear, coincides with the exact result at $T=0$
\cite{Langer:1962a:PR,Sharapov:2003:PRB}.  Even though this
procedure allows us to correctly reproduce the optical
conductivity in the low-frequency limit, it does not solve the
problem of knowing {\em a priori} the sum rule corresponding to
this approximated optical conductivity. Indeed, since this
symmetric bubble is not exact for the full model
\pref{Hamiltonian}, and in contrast to the bubble \pref{pigi}
contains two full vertices, one cannot expect that the optical
conductivity calculated using this bubble would satisfy the usual
sum rule corresponding to the diamagnetic tensors \pref{deft} or
\pref{deftmf}.

However, this last issue can be solved exactly by applying the same GI
arguments to the reduced quadratic model \pref{DDW-Hamiltonian}.  Indeed,
the static limit $\G_i\equiv \g_i+V_i({\bk})$ of the full vertex function,
obtained from the original interacting model, can also be considered as a
bare vertex $\tilde{\gamma}$ for the DDW Hamiltonian
\pref{DDW-Hamiltonian}. Moreover, since this Hamiltonian describes
noninteracting quasiparticles, the bare and the full vertex coincide, so
that the symmetric correlator $\Pi^{DDW}$ in Eq.~\pref{pigi-DDW} is the
{\em exact} one for this model. As a consequence, the diamagnetic term
\pref{new.tensor} of the reduced model gives the sum rule $W^{DDW}(D,T)$
\pref{sum.T>Tc} for the symmetric current operator \pref{pigi-DDW}, which
is the exact one within the quadratic theory \pref{DDW-Hamiltonian} and at
the same time provides us with a good approximation for the optical
conductivity of the true interacting system, at least at low energy.

The last issue we addressed in the present paper is to analyze to which
extent the sum rule \pref{sum.T>Tc} can be related to the behavior of the
microscopic Hamiltonian \pref{Hamiltonian}. In general, it is believed that
in the presence of interactions the restricted sum rule \pref{weight.band},
derived for the electrons within the lowest conducting tight-binding band,
is still valid, provided that the occupation number $n_{\bk \s}$ takes into
account the effect of the interactions. In this case, we should rely on the
estimate $W(D,T)$ in Eq.~\pref{deftmf} for the sum rule in the DDW
state. However, this approach has two disadvantages: (i) we cannot derive
the optical-conductivity which would lead to this sum rule; (ii) no general
argument holds to justify why this attitude is the correct one to estimate,
at mean-field level, the sum rule for the interacting microscopic model.
Motived by these observations we argued that in the case of interactions
leading to a DDW formation a better mean-field approach to the transport
properties is provided by the calculation of the optical conductivity by
means of the bubbles $\Pi^{DDW}$. Thus, to obtain the correct mean-field
approximation for the spectral-weight behavior is not sufficient to modify
the occupation number $n_{\bk\s}$ below $T_{DDW}$, but it is more likely
that a proper re-definition of the diamagnetic tensor is needed. As a
consequence, the sum rule should be estimated by means of Eq.
\pref{sum.T>Tc} instead of Eq.~\pref{deftmf}, leading to an {\em increase}
of spectral weight below $T_{DDW}$. However, this assumption would require
also that the integrated spectral weight $N^{DDW}(\o)$ in the DDW state
becomes larger than the $N(\o)$ for the non-interacting system at some
``low'' frequency $\o_M$. As we discussed in
Sec.~\ref{sec:conductivity-DDW}, such cut-off frequency $\o_M$ turns out to
be in general lower than the plasma edge, but its precise value depends
crucially on the parameters of the DDW transition (doping, order parameter
at $T=0$, etc.). Moreover, it is not clear yet if a $\o_M$ below the plasma
frequency is a sufficiently low-energy scale for the interacting
tight-binding model, since no universal definition exists of the frequency
itself below which the restricted sum rule should be applicable.

Finally, a comment is in order now about the comparison between our results
and the experimental optical data for cuprates. A first issue is related to
the fact that the most recent experiments show that the temperature
variation of the spectral weight is larger than expected in a tight-binding
estimate also in optimally doped and overdoped compounds
\cite{Santander:2004,Ortolani:2004} (see also Appendix A), i.e. eventually
at doping larger than the critical doping $\d_0=0.2$ for the charge
ordering phenomenon. This could mean that a more general effect of the
strong correlations present in these materials can be responsible for the
large temperature variation of $W(T)$. This possibility has been indeed
investigated recently in Ref. \cite{Toschi:2004}, where $W(T)$ has been
evaluated by means of the DMFT (dynamical mean field theory) approach to
the Hubbard model, which seems to reproduce the large temperature
variations of $W(T)$ observed in the experiments.  A second
issue arises about the lack, in the experiments, of a clear signature of an
inter-band conductivity as the one reported in Fig. 4. In particular, BSCCO
compounds, that were used as a paradigm for the choice of DDW parameter
values in cuprates \cite{Benfatto:2000:EPJ}, exhibit in general a quite
featureless conductivity, with a slowly decaying high-frequency
tail. However, in different families of cuprates, displaying a similar
spectral-weight behavior, clear signatures of charge ordering have been
indeed observed in the optical spectra, even though located at much lower
energy scales with respect to the one obtained here using the parameter
values for BSCCO compounds.  This is the case of LSCO and YBCO, where
far-infrared features, well separated from the Drude peak, have been
measured recently \cite{stripes}.  In both references
\cite{caprara,stripes2} these features were actually interpreted as due to
a charge-ordering phenomenon, described by means of some different
theoretical approaches which did not allow one to discuss at the same time
the issue of the spectral-weight behavior.  For these reasons, even though
the analysis presented here cannot be conclusive as far as the optical
spectra of HTSC are concerned, we believe that a deeper investigation of
the role of charge degrees of freedom can eventually lead to a better
understanding of the conductivity of cuprates. At the same time, the
analysis presented here could be extended to other systems like
2$H$-TaSe$_2$, where a $k$-space modulated CDW forms \cite{Neto:2001:PRL}
and where clear signatures of a Drude response accompanied by a
mid-infrared peak have been observed in the optical spectra
\cite{Vescoli:1998:PRL}.

\section{Acknowledgments}
We are grateful V.P.~Gusynin for bringing to our attention the derivation
of symmetrized expression for the derivatives of the polarization operator.
We acknowledge  stimulating discussions with A.~Toschi, M.~Ortolani and
P.~Calvani. This work was supported by the research projects
2000-067853.02/1, 620-62868.00, and MaNEP (9,10,18) of the Swiss National
Science Fundation.

\appendix

\section{Failure of the tight-binding estimate}
\lb{sec:appendix_A}

As we noticed in the Introduction, even though a $T^2$ temperature decrease
of $W(T)$ is observed in the experiments, the measured slope is quite
larger than the one expected within the simple non-interacting
tight-binding estimate \pref{weight.normal}. To quantify this discrepancy
in the most accurate way, we evaluate explicitly the spectral weight for
the tight-binding model by including also a next-nearest neighbors term in
the band dispersion, see Eq. \pref{etp}. As a consequence, $W(T)$ is given
by Eq. \pref{wntp}. 

\begin{figure}[htb]
\centering{ \includegraphics[width=7.cm,angle=-90]{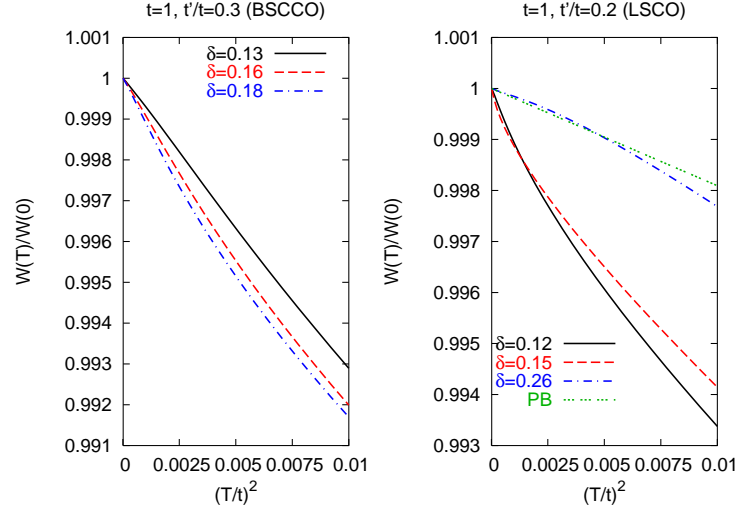} }
\caption{(Color online) Spectral weight $W(T)/W(0)$ according to
Eq.~\pref{wntp} for $t'=0.3t$ (left panel, appropriate for BSCCO
\cite{Santander:2002:PRL,Santander:2004}) and $t'=0.2t$ (right panel,
appropriate for LSCO \cite{Ortolani:2004}) at various doping. In the right
panel we also report the result obtained with the estimate
\pref{weight.normal} and the parabolic band (PB in the caption)
approximation for $c(\mu)=1/4\pi t$, using the value $W(0)=0.682t$ at
$\d=0.26$.}
\label{fig:7}
\end{figure}

To correctly reproduce the Fermi surface of BSCCO and LSCO compounds we
will assume $t=0.3 eV$ and $t'=rt$, where $r=0.3$ for BSCCO and $r=0.2$ for
LSCO (where the Fermi surface changes topology in the overdoped region,
becoming electron-like at about $\d=0.2$ doping \cite{Ino:2002:PRB}). The
results of $W(T)/W(0)$ as a function of $(T/t)^2$ from Eq.~\pref{wntp} are
reported in Fig. 7 for several doping (by fixing as usual the correct
chemical potential at each doping and temperature from the self-consistency
equation for the particle number). In the left panel we report the
estimate for BSCCO, that should be compared to the experimental data (for
underdoped, optimally doped and overdoped samples) of Ref.
\cite{Molegraaf:2002:Science,Santander:2002:PRL,Santander:2004}. Observe
that in Ref.\cite{Santander:2002:PRL,Santander:2004} the variation of
$W(T)/W(0)$ between room temperature and $T=0$ of the order of $20 \% -5
\%$ when measured at various cut-off frequencies, while the tight-binding
estimate in Fig. 7 never exceed the $0.6 \%$ (for $t=0.3$ eV $T=300$ K
corresponds to $(T/t)^2=0.0074$). Analogous considerations hold for the
comparison between the measured spectral weight in LSCO
\cite{Ortolani:2004} and the estimate \pref{wntp} reported in the right
panel of Fig. 7. A comment is in order now about the role of the Van Hove
singularity (VHS) in the density of states. Indeed, according to
Eq.~\pref{weight.normal}, where the $t'=0$ case was considered, the
coefficient $c(\mu)=\mu N'(\mu)+N(\mu)$ could increase considerably by 
approaching the VHS. This effect is indeed seen in the curves at $\d=0.12$
and $\d=0.16$, where the initial slope of $W(T)$ is quite large. However,
as soon as the temperature increases the effect of the VHS is washed out
very rapidly and the overall variation in the range of $T$ between $0-0.1t$
attains the same values found for the case $t'=0.3t$. Moreover, for the
overdoped case $\delta=0.26$ the slope of $W(T)/W(0)$ agrees very well with
the approximation $c(\mu)=1/4\pi t$ of the parabolic band dispersion (PB in
the figure), which would give the value $\pi^2 c(\mu)/6=0.13/t$ for the
coefficient in Eq.~\pref{weight.normal}. For these reasons one can conclude
that the $t'$ term in the band dispersion \pref{etp} has a minor role in
determining the spectral-weight variations, and indeed it was only briefly
discussed in Sec. IV C of the present work.

\section{Sum rule in the presence of disorder}
\lb{sec:appendix_B}

As we did in Sec.~\ref{sec:conductivity-DDW} we will take into account the
effect of disorder by using the substitution \pref{dis} in the spectral
representation \pref{spectral} of the Green's function. To see how
Eq. \pref{sc2} is modified we rewrites it in terms if the spectral
function:
\be
n=\frac{2T}{N}\sum_{\bk,i\o_n}^{RBZ} Tr[G(\bk,i\o_n)]e^{i\o_n0^+}=
\frac{2}{N}\sum_{\bk}^{RBZ}\int dz [M(z-E)+M(z+E)]f(z-\mu).
\ee

Analogously, the definition \pref{new.tensor} of the diamagnetic tensor in
the DDW state can be expressed as:
\bea
\langle \tau_{ii} \rangle
&=&-\frac{T}{N}\sum_{\bk,i\o_n}^{RBZ} \left\{\e_\bk
Tr[G(\bk,i\o_n)\s_3]-D_\bk Tr[G(\bk,i\o_n)\s_2]\right\}e^{i\o_n0^+}=\nn\\
&=&-\frac{1}{N}\sum_{\bk}^{RBZ}E_\bk
\int dz [M(z-E)-M(z+E)]f(z-\mu).
\eea
Observe that if one puts $M(z)=\delta(z)$ the results \pref{sc2} and
\pref{sum.T>Tc} can be recovered, and for $D_0=0$ one finds the
corresponding expressions for the normal state. At T=0, which is the case
considered in Fig. 5, the previous equations simplify. Indeed, since the
Fermi functions reduce to a step function, one has:
\be
\int_{-\infty}^{\infty} dz M(z-E)f(z-\mu)=
\int_{-\infty}^{\mu} dz \frac{1}{\pi}\frac{w}{z^2+w^2}=\frac{1}{\pi}\left(
\arctan\frac{\mu-E}{w}+\frac{\pi}{2}\right)
\ee
so that the self-consistency equation for the particle number and the
equation for the spectral weight can be written as:
\bea
n-1=\frac{2}{\pi N}\sum_{\bk}^{RBZ} \left(
\arctan\frac{\mu-E}{w}+\arctan\frac{\mu+E}{w} \right),\\
\frac{W^{DDW}(D,T=0)}{(\pi e^2 a^2/V)}=
-\frac{1}{\pi N}\sum_{\bk}^{RBZ}E_\bk  \left(
\arctan\frac{\mu-E}{w}-\arctan\frac{\mu+E}{w} \right).
\eea
Observe that here we did not consider the effect of the DDW formation on
the transport scattering time, which can be present. For a detailed
discussion of impurity scattering in the DDW state see
Ref. \cite{Zhu:2001:PRL}.


\begin{thebibliography}{99}
\bibitem{footnote1}
The definition \pref{vbare} for a lattice system
is correct only in the limit of small $\bq$, when
one can approximate $\e_{\bk+\bq}-\e_{\bk}\approx \bq \cdot
v_\bk^F$. However, since in the following we are interested in discussing
the optical conductivity, which is related to the $\bq=0$ limit of the
current operator \pref{defj}, we can adopt the notation
\pref{defj}-\pref{vbare} without loss of generality.
\bibitem{noteCDW} The Hamiltonian \pref{DDW-Hamiltonian} does not display
any explicit charge modulation, even though a gap appears in the
quasiparticle spectrum. To describe a true CDW modulation a coupling like
$|D_\bk|c^+_{\bk\s}c_{\bk+\bQ\s}+h.c.$ should be considered instead.
\end{thebibliography}
\end{document}